\documentclass[lettersize,journal]{IEEEtran}
\usepackage{amsmath,amsfonts}
\usepackage{algorithmic}
\usepackage{algorithm}
\usepackage{array}

\usepackage{textcomp}
\usepackage{stfloats}
\usepackage{url}
\usepackage{verbatim}
\usepackage{graphicx}
\usepackage{cite}
\usepackage{amssymb}
\usepackage{amsfonts} 
\usepackage{physics} 
\usepackage{graphicx}
\usepackage{color} 
\usepackage{bm}
\usepackage{epstopdf} 
\usepackage{subfigure}
\usepackage{placeins}
\usepackage{booktabs}  
\usepackage[dvipsnames]{xcolor} 
\usepackage{cite} 
\usepackage{balance}
\usepackage{arydshln} 
\usepackage{mathrsfs} 
\usepackage{cuted} 

\usepackage{multirow} 
\usepackage{makecell} 
\usepackage{hyperref}
\usepackage{tabularray}
\usepackage{multirow}

\newcommand{\figref}[1]{Fig. \ref{#1}}

\DeclareMathAlphabet{\mathsfit}{\encodingdefault}{\sfdefault}{m}{sl}
\SetMathAlphabet{\mathsfit}{bold}{\encodingdefault}{\sfdefault}{bx}{n}
\newcommand{\tens}[1]{\bm{\mathsfit{#1}}}
\def\tA{{\tens{A}}}
\def\tB{{\tens{B}}}
\def\tC{{\tens{C}}}
\def\tD{{\tens{D}}}

\def\tG{{\tens{G}}}
\def\tH{{\tens{H}}}

\def\tP{{\tens{P}}}

\def\tX{{\tens{X}}}

\begin{document}
	
\title{Deep Learning-Based Joint Uplink-Downlink\\ CSI Acquisition for Next-Generation\\ Upper Mid-Band Systems}
	
\author{Xuan He, \textit{Graduate Student Member}, \textit{IEEE}, Hongwei Hou,  \textit{Graduate Student Member}, \textit{IEEE},\\ Yafei Wang,  \textit{Graduate Student Member}, \textit{IEEE}, Wenjin Wang, \textit{Member}, \textit{IEEE}, Shi Jin,  \textit{Fellow}, \textit{IEEE}, \\
Symeon Chatzinotas,  \textit{Fellow}, \textit{IEEE}, Bj{\"o}rn Ottersten,  \textit{Fellow}, \textit{IEEE}
		\thanks{Xuan He, Hongwei Hou, Yafei Wang and Wenjin Wang are with the National Mobile Communications Research Laboratory, Southeast University, Nanjing 210096, China, and also with Purple Mountain Laboratories, Nanjing 211100, China (e-mail: hexuan0502@seu.edu.cn; hongweihou@seu.edu.cn; wangyf@seu.edu.cn;  wangwj@seu.edu.cn).}
		\thanks{Shi Jin is with the National Mobile Communications Research Laboratory, Southeast University, Nanjing 210096, China (e-mail: jinshi@seu.edu.cn).}
        \thanks{Symeon Chatzinotas and Bj{\"o}rn Ottersten are with the Interdisciplinary Centre for Security, Reliability and Trust (SnT), University of Luxembourg (E-mails: \{symeon.chatzinotas, bjorn.ottersten\}@uni.lu).}
	}
	
	\markboth{}
	{}
	%
	
	\maketitle
	\begin{abstract}
    In next-generation wireless communication systems, the newly designated upper mid-band has attracted considerable attention, also called frequency range 3 (FR3), highlighting the need for downlink (DL) transmission design, which fundamentally relies on accurate channel state information (CSI). However, CSI acquisition in FR3  systems faces significant challenges: the increased number of antennas and wider transmission bandwidth introduces prohibitive training overhead with traditional estimation approaches, as each probing captures only incomplete spatial-frequency observation, while higher carrier frequencies lead to faster temporal channel variation. 
    To address these challenges, we propose a novel CSI acquisition framework  that integrates CSI feedback, uplink (UL) and DL channel estimation, as well as channel prediction in the FR3 time division duplex (TDD) massive multi-input multi-output (MIMO) systems. 
    Specifically, we first develop the Joint UL and DL Channel Estimation Network (JUDCEN) to fuse incomplete observations based on the sounding reference signals (SRSs) and CSI-reference signals (CSI-RSs). 
    By exploiting the complementary characteristics of preliminary UL and DL estimation features, obtained through initial UL estimation and quantized-feedback-assisted DL estimation,  it enables full CSI reconstruction in the spatial domain.
    To mitigate the performance degradation in the feedback process, we propose the Transformer-MLP CSI Feedback Network (TMCFN),  employing an MLP-based module to jointly exploit angle- and delay-domain features.
    Building upon the reconstructed full CSI, we further develop the Mamba-based Channel Prediction Network (MCPN), which exploits selective state-space model (SSM) mechanism to capture long-range temporal dynamics in the angle-delay domain for future CSI prediction.
    Simulation results demonstrate that the proposed framework consistently outperforms benchmarks in  both CSI acquisition accuracy and transmission spectral efficiency with lower computational complexity.
	\end{abstract}
	\begin{IEEEkeywords}
		Joint uplink-downlink transmission, deep learning, FR3, channel estimation and prediction.
	\end{IEEEkeywords}

	%
\IEEEpeerreviewmaketitle

\bstctlcite{IEEEexample:BSTcontrol}
	
\section{Introduction}\label{Section 1}
\IEEEPARstart{A}{s} 
 the sixth-generation (6G) communication systems continue to evolve, a variety of promising applications are emerging, such as extended reality and holographic communication, which impose stringent requirements on both the coverage and capacity of future wireless networks \cite{giordani2020toward}. 
However, the existing frequency ranges (FRs), including FR1 (sub-6 GHz) and FR2 (millimeter-wave), are challenged by spectrum scarcity and severe path loss, respectively, driving the exploration of new spectrum to achieve a more favorable trade-off between coverage and capacity \cite{naqvi20215g}.
 In this context, the newly designated upper mid-band (FR3), spanning 7-24 GHz, has attracted growing attention as a promising compromise, offering an appealing balance between coverage performance and available bandwidth \cite{2022Samsung, 2022je, 2023WRC, hou2025tensor}. 
 
 Accordingly, reliable multi-user downlink (DL) transmission in FR3 time division duplex (TDD) systems is of paramount importance, which fundamentally depends on the accuracy of channel state information (CSI) acquisition \cite{wang2022weighted,wang2025statistical,11049893, WANG2025}.
 Nevertheless, the unique propagation characteristics of FR3 present new challenges for CSI acquisition-most notably, the growing gap between limited pilot resources and the increased probing demands in the spatial, frequency, and temporal domains.
 Specifically, the short wavelength enables denser antenna integration with a given physical aperture, making full spatial domain probing infeasible under pilot resource constraints in the current specifications \cite{3gpp.38.211}, while the large bandwidth introduces a similar issue in the frequency domain.
 In the temporal domain, the higher carrier frequency of FR3 compared to FR1, along with richer multipath propagation than FR2 leads to more rapid channel variations \cite{miao2023sub}. 
 Therefore, these unique spatial, frequency, and temporal domain characteristics of FR3 collectively underscore the need for effective and scalable CSI acquisition strategies.
 
\subsection{Previous Work}
In wireless communication systems, CSI acquisition is commonly carried out via two main approaches: uplink (UL) training and DL training \cite{zheng2022survey}. 
For UL training approaches, the base station (BS) employs UL reference signals, specifically the sounding reference signal (SRS) as per existing specifications, to estimate UL CSI and then derives DL CSI based on the reciprocity between UL and DL channels \cite{li2000optimum,noh2006low,zhu2024joint,dong2019deep,lin2021deep}. In the earliest UL training attempts, conventional least-squares (LS) estimation is adopted to obtain the spatial-frequency   channel \cite{li2000optimum}, which is further refined via the linear minimum mean square error (LMMSE) method \cite{noh2006low}. 
Considering computational complexity and limited transmission ports, the BS obtains only partial spatial-frequency domain channel observations, motivating the development of neural network (NN)-based channel extrapolation methods.
In \cite{dong2019deep}, a convolutional neural network (CNN)-based estimator is introduced to process the spatial-frequency domain channel of adjacent subcarriers. For the spatial domain, the fully-connected (FC) neural network is proposed to infer the unobserved CSI from that available on a subset of antennas  \cite{lin2021deep}.  

In contrast to UL CSI acquisition, the DL training is constrained by limited transmit ports, as well as pilot resources and feedback overhead that scale with the number of BS antennas. 
To alleviate these limitations, compressed sensing methods \cite{lee2016channel, wu2021hybrid,hou2024beam} have been proposed to exploit the channel sparsity of massive multi-input multi-output (MIMO) operating in mmWave bands, transforming CSI acquisition into an estimation problem of reduced physical parameters, such as angle of departure (AoDs) and angle of arrival (AoAs). However, these methods are  sensitive to practical propagation environments and antenna array calibration \cite{fang2023multi}, motivating the adoption of more robust NN–based approaches. Specifically, in \cite{zhou2023pay}, a dual-attention-based channel estimation network (DACEN) is proposed to jointly learn the spatial-delay domain features for DL channel estimation under low-density pilots. 
However, most of the aforementioned works either assume perfect feedback or neglect CSI feedback, whereas limited feedback remains a fundamental bottleneck in DL-based CSI acquisition methods \cite{zhuang2025extract}. The first deep learning-based CSI feedback network, CsiNet \cite{wen2018deep} demonstrated significant performance gains over conventional methods. Building upon this, TransNet \cite{cui2022transnet} integrated  the Transformer architecture \cite{vaswani2017attention} to further enhance CSI feedback accuracy. 

Additionally, the higher carrier frequency of FR3 leads to faster temporal channel variation, necessitating channel prediction or time-domain channel extrapolation \cite{hou2024tensor}. Model-based approaches such as the sum-of-sinusoids (SOS) method \cite{wong2006wlc43} and statistical prediction methods including autoregressive (AR) models \cite{duel2007fading, wu2021channel, lv2019channel} have shown promise in channel prediction. Recent advances in deep learning have also significantly enhanced channel prediction capabilities, offering a promising solution \cite{he2025sca}.  Specifically, a temporal UL and DL channel extrapolation network (TUDCEN) empowered  by generative artificial intelligence (AI) is proposed in \cite{zhou2025low} for slot-level channel extrapolation.

However, the increased number of antennas and wider bandwidth of the FR3 band render the standalone UL or DL pilot resources insufficient to probe the full spatial-frequency channel, making joint UL and DL-based methods more promising. 
In \cite{salim2011combining}, the idea of combining UL pilot training with DL quantized feedback is first introduced in long term evolution (LTE) TDD systems. 
Moreover, the ratio technique and angle-domain pre-search technique are employed in the single-carrier system \cite{lee2021downlink}, which extract information from quantized DL CSI-RS feedback to assist SRS-based UL channel estimation. 
Nevertheless, the joint exploitation of UL and DL signals for spatial–frequency channel reconstruction still remains nontrivial, particularly in MIMO-OFDM systems where CSI-RS and SRS pilot allocations are more intricate.

\subsection{Motivations and Contributions}
Reliable DL transmission hinges on accurate CSI acquisition, while both UL and DL approaches in FR3 systems suffer from fundamental challenges across the spatial, frequency, and temporal domains \cite{3gpp.38.211,miao2023sub}.
Despite increasing research efforts, existing solutions suffer from several critical limitations, such as inefficient exploitation of CSI-RS resources and incompatibility  with current specifications.

Motivated by prior work, this paper investigates the integrated design of CSI feedback, UL and DL channel estimation, and channel prediction. To the best of our knowledge, this is the first work that jointly exploits UL and DL signals for spatial-frequency domain channel reconstruction and prediction in massive MIMO-OFDM TDD FR3 systems. The main contributions of this paper are summarized as follows:
	\begin{itemize}
		\item For multi-domain full CSI acquisition, we propose a joint UL and DL transmission framework that integrates UL pilot training and DL quantized feedback. 
        Specifically, the proposed framework obtains preliminary channel features through UL and DL initial channel estimation and enables the BS to collect incomplete UL and DL observations via quantization feedback. 
        Due to the complementary observation structure of CSI-RSs and SRSs, the full CSI in the spatial-frequency domain can be reconstructed by cross-referencing, providing a foundation for subsequent channel prediction.
		\item To achieve more accurate DL CSI reconstruction at the BS, we develop the Transformer-MLP-based CSI Feedback Network (TMCFN) within the proposed framework. Specifically, TMCFN upgrades the feedforward layer (FFL) of traditional Transformers by a multi-layer perceptron (MLP)-based module to simultaneously leverage sparsity in both the angle and delay domains. 
        With the collected UL and DL CSI at the BS, we propose a Joint UL and DL Channel Estimation Network (JUDCEN). The JUDCEN fuses UL and DL CSI in the spatial-frequency domain through spatial-frequency attention mechanisms, where convolution and pooling operations yield an attention map to guide the feature mapping.
		\item To address the rapid temporal channel variations, we design the Mamba-based Channel Prediction Network (MCPN). The MCPN is built on stacked Mamba modules, which achieves long-sequence temporal dependency capture in the angle-delay domain by leveraging the selective state space model (SSM) mechanism.
        The TMCFN, JUDCEN and MCPN collectively constitute the Hybrid Attention- and SSM-Based CSI Acquisition Network (HASCAN), which enables accurate CSI reconstruction tailored to the unique reference signal structure of the FR3 system within the proposed framework. 
        To address convergence issues caused by the increased model complexity, we adopt a structured training strategy comprising pre-training, parameter pre-loading, and joint fine-tuning. 
        The conventional MSE loss is replaced by a cosine similarity-based loss, which better captures directional characteristics of CSI and leads to improved spectral efficiency, while maintaining comparable computational complexity. Simulation results show that, the proposed HASCAN  consistently outperforms benchmarks in both CSI acquisition accuracy and spectral efficiency by jointly leveraging UL and DL signals.
	\end{itemize}
	
	This paper is structured as follows: Section \ref{Section 2} introduces the system model and conventional CSI acquisition approaches. In Section \ref{Section 3}, the proposed framework is derived under practical system configurations. The HASCAN is proposed in \ref{Section 4}, and simulation results are given in Section \ref{Section 5}. Finally, we conclude this work in Section \ref{Section 6}.
	
	\emph{Notation:} $\left(\cdot\right)^T$ and $\left(\cdot\right)^H$ denote the 
	transpose and conjugate transpose operators, respectively. $x$, $\mathbf{x}$, $\mathbf{X}$, $\tX$ respectively denote a scalar, column vector, matrix, and tensor. $\mathbf{F}_{G} $ denotes the $G\times G$ discrete Fourier transform (DFT) matrix. The column, row, and tube fibers of a third-order tensors $\tX$  are denoted by $\mathbf{x}_{: jk}$, $\mathbf{x}_{i: k}$, and $\mathbf{x}_{ij:}$, respectively. $\bar{\jmath}=\sqrt{-1}$ denotes the imaginary unit.  $[\cdot]_{i}$ and $[\cdot]_{i,j}$ denote the $i$-th element of a vector and the $(i,j)$-th element of a matrix. $ \otimes $ and $\odot$ denote the Kronecker and Hadamard product, respectively.  $\left\|\cdot\right\|_2  $ denotes L2 norm.  $ \delta\left(\cdot\right) $ and $ \sigma\left(\cdot\right) $ denote the Dirac delta and sigmoid functions, respectively. $ \operatorname{diag}\left\{\mathbf{x}\right\} $ denotes a diagonal matrix with main diagonal $\mathbf{x}$. $ \operatorname{E}\left(\cdot\right) $, $ \operatorname{Conv}\left(\cdot\right) $, $ \operatorname{AvgP}\left(\cdot\right) $, and $ \operatorname{MaxP}\left(\cdot\right) $ denote expectation, convolutional, average-pooling, and max-pooling operations, respectively. 
	\section{System Model and Problem Formulation }\label{Section 2}
    
    In this paper, we consider a massive MIMO-OFDM system operating in TDD mode, where the BS is equipped with $ N_{\rm Tx}$  antennas  to serve multiple user equipment (UE) each  with $ N_{\rm Rx}$ antennas, and $  N_{\rm c} $ subcarriers are adopted for data transmission. 
    
    \subsection{Channel Model}
        \begin{figure}[tp!] 
		 	\centering  
		 	\subfigure[]{
                \includegraphics[width= 0.6\linewidth]{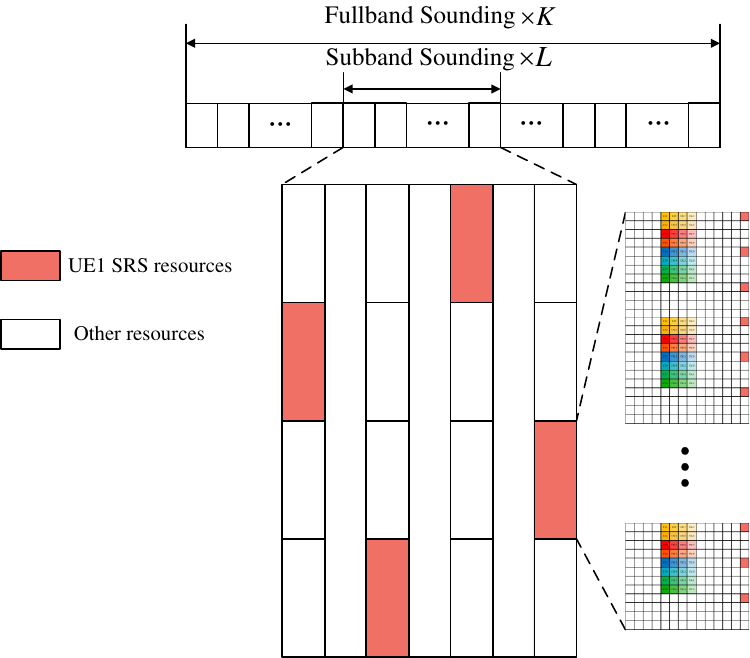}\label{systema}
            }\\
		      \subfigure[]{
		 		\includegraphics[width = 0.6\linewidth]{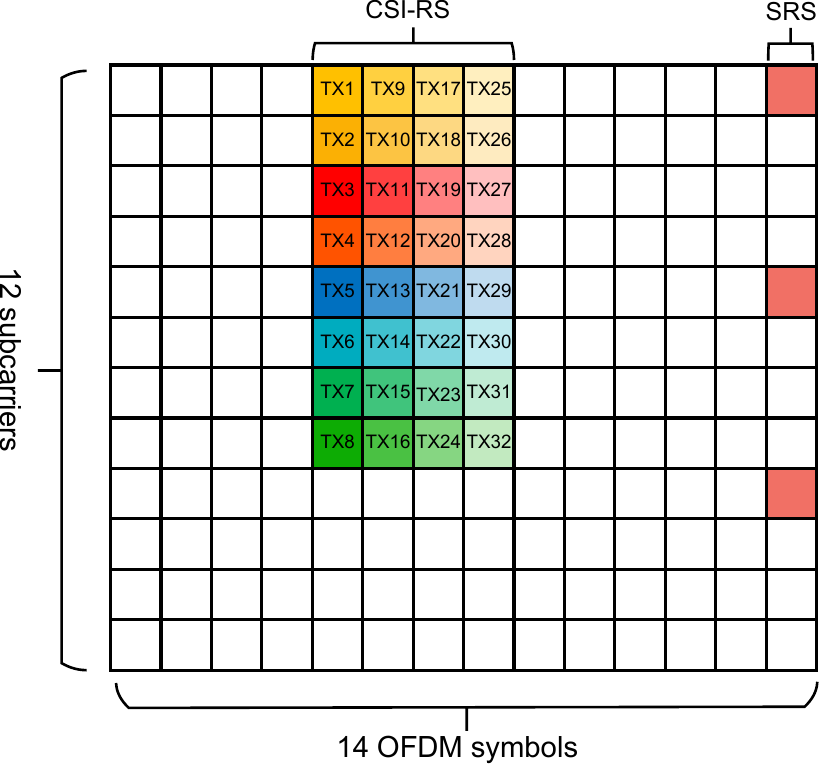}\label{systemb}
            }
		\caption{The typical CSI-RS and SRS pattern (a) within the full band; (b) within one RB. \cite{3gpp.38.211}}
        \vspace{-5mm}
		 \end{figure}
With the perfect antenna array calibration assumption \cite{fang2023multi}, the time-varying spatial domain channel impulse response (CIR) matrix $\mathbf{H}(t, \tau)\in \mathbb{C}^{N_{\rm Tx}\times N_{\rm Rx}}$ can be expressed by \cite{hou2024joint}
\begin{align}
    \mathbf{H}(t, \tau)=\sum_{p=0}^{P-1} \alpha_p e^{\bar{\jmath} 2 \pi \nu_p t}  \delta \left(\tau-\tau_p\right)\mathbf{a}_{\rm T}\left(\theta_{{\rm T},p}\right)\mathbf{a}_{\rm R}^H\left(\theta_{{\rm R},p}\right),\label{time-varying}
\end{align}
where $\alpha_p$, $\tau_p$, $\nu_p$, $\theta_{{\rm T},p}$, and $\theta_{{\rm R},p}$ denote the complex gain, propagation delay, Doppler frequency, AoD, and AoA of the $p$-th path, respectively, $P$ is the number of propagation paths between the typical UE and BS, and $\mathbf{a}_{\rm T}\left(\cdot\right)$ and $\mathbf{a}_{\rm R}\left(\cdot\right)$ denote the angle domain steering vectors \cite{hou2024tensor} for the BS and UE, respectively.  

By taking the Fourier transformation of CIRs, the time-varying spatial domain channel frequency response $\mathbf{H}(t, f)\in \mathbb{C}^{N_{\rm Tx}\times N_{\rm Rx}}$ at time $t$ and frequency $f$ can be expressed as
\begin{align}
    \mathbf{H}(t, f)=\sum_{p=0}^{P-1} \alpha_p e^{\bar{\jmath} 2 \pi \nu_p t}e^{-\bar{\jmath} 2 \pi \tau_p f}  \mathbf{a}_{T}\left(\theta_{{\rm T},p}\right)\mathbf{a}_{R}^H\left(\theta_{{\rm R},p}\right)\label{time-varyingsf}.
\end{align}
Furthermore, we define the spatial domain channel sampled at discrete subcarriers and slots as 
\begin{align}
     \mathbf{H}\{n,k\}=\mathbf{H}(n\Delta t, k\Delta f)\in \mathbb{C}^{N_{\rm Tx}\times N_{\rm Rx}},\label{H_d}
\end{align}
where  $\Delta t$ denotes time interval between two adjacent slots with CSI-RS allocated, the time interval between two SRS transmissions is assumed to be  $S\Delta t$,  and $\Delta f$ denotes the subcarrier spacing.
\subsection{Unique Characteristics of the FR3 Channel}
CSI acquisition in wireless communication systems generally relies on UL or DL pilot-based estimation. However, the unique characteristics of FR3 systems, such as more antennas at both BS and UEs, wider transmission bandwidth, and faster time-variation, introduce new challenges to the traditional pilot-based estimation schemes.
\subsubsection{More Antennas}
As carrier frequencies increase in FR3, more antennas can be packed within a fixed physical aperture at both the BS and UEs. 
However, CSI-RS and SRS transmissions are constrained by the maximum number of supported ports, i.e., $M^{\rm max}_{\rm c}=32< N_{\rm Tx} $ and $M^{\rm max}_{\rm s}=4 < N_{\rm Rx}$,  defined by the 3rd generation partnership project (3GPP) specifications \cite{3gpp.38.211}, leading to incomplete spatial observations at both ends, as shown in \figref{diff_csirs_srs_v1}.
\subsubsection{Wider Transmission Bandwidth}
In FR3 systems, the substantial increase in system bandwidth makes FHS mode a mandatory choice for SRS transmission, due to constraints on transmission power and radio frequency (RF) operating bandwidth \cite{zhu2024joint}. However, SRS in FHS mode provides only partial and time-staggered observations in the frequency domain.

\subsubsection{Faster Time-Variation} 
The elevated carrier frequency in FR3 systems leads to higher Doppler frequencies for a given mobility profile, resulting in rapid temporal variations of the CSI \cite{miao2023sub}. Therefore, CSI obtained from either CSI-RS or SRS becomes outdated for data transmission, which is further exacerbated by the spatial-frequency domain partial observation structure, posing significant challenges for accurate and consistent channel prediction.
 	\begin{figure}[t]
		\centering
		\includegraphics[width=3in]{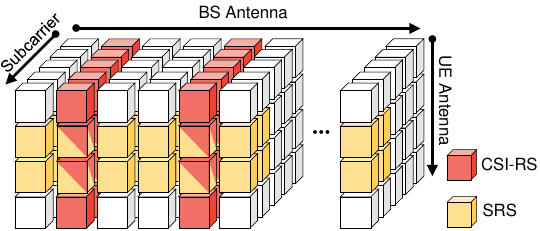} 
		\caption{Different structure of uplink and downlink reference signals.} 
		\label{diff_csirs_srs_v1} 
        \vspace{-5mm}
	\end{figure}
\subsection{Conventional CSI Acquisition Approaches in FR3 Systems}
We consider two practical CSI acquisition approaches in the 3GPP specifications: (i) UE-side acquisition based on CSI-RS followed by quantized feedback, and (ii) BS-side acquisition based on SRS and channel reciprocity. Since CSI-RS is the same for all UEs and SRS follows a comb-type pattern, 
inter-UE interference can be neglected; thus, UE indices are omitted for clarity. In the following, we review the received signal models and highlight the key principles for each strategy.
\subsubsection{CSI Acquisition Based on CSI-RS}
For the typical CSI-RS configuration with $M_{\rm c}$ ports, as illustrated in Fig. \ref{systemb}, we assume that CSI-RS resources are continuously allocated across $N_{\rm RB}$ resource blocks (RBs) and uniformly distributed over $N_{\rm dt}$ slots in the time domain during DL training.
Therefore, the received signal of CSI-RS corresponding to the $s$-th slot and the $r$-th RB can be expressed by 
\begin{align}
	\mathbf{Y}^{\rm d}_{s,r} = \mathbf{X}^{\rm d}\mathbf{F}^{\rm d}_{s} \mathbf{H}^{\rm d}_{s,r} + \mathbf{N}^{\rm d}_{s,r}\in \mathbb{C}^{M_{\rm c}\times N_{\rm Rx}},
    \label{dltraining}
\end{align}
where $\mathbf{H}^{\rm d}_{s, r} = \left[  \mathbf{H}\{s,rM_{sc}+1\}^{T}, \dots, \mathbf{H}\{s,(r+1)M_{\rm sc}\}^{T}  \right]^{T}$
$\in \mathbb{C}^{M_{\rm sc}N_{\rm Tx}\times N_{\rm Rx}}$ denotes the spatial-frequency domain channel for the $s$-th slot and the $r$-th RB, $M_{\rm sc}$ denotes the number of subcarriers within one RB, $ \mathbf{X}^{\rm d}\in \mathbb{C}^{M_{\rm c}\times M_{\rm c}} $, $ \mathbf{F}^{\rm d}_{s}\in \mathbb{C}^{M_{\rm c}\times M_{\rm sc}N_{\rm Tx}} $, and $ \mathbf{N}^{\rm d}_{s, r}\in \mathbb{C}^{M_{\rm c}\times N_{\rm Rx}} $ denote the CSI-RS pilot signal, spatial filter, and noise, respectively. Due to pilot resource constraint in FR3 systems $  M_{\rm c}<N_{\rm Tx} $, the UE can only obtain the estimation of the low-dimensional equivalent DL channels $ \mathbf{G}^{\rm d}_{s,r} = \mathbf{F}^{\rm d}_{s} \mathbf{H}^{\rm d}_{s,r} \in \mathbb{C}^{M_{\rm c}\times N_{\rm Rx}}$ for feedback. By aggregating the received signals across all RBs, the full-band equivalent spatial-frequency domain channel  $\mathbf{G}^{\rm d}_{s} = [(\mathbf{G}^{\rm d}_{s, 1})^{T}, \dots, (\mathbf{G}^{\rm d}_{s, N_\text{RB}})^{T}]^{T} \in \mathbb{C}^{N_\text{RB}M_{\rm c}\times N_{\rm Rx}}$ can be reconstructed, which is then compressed and quantized into a feedback vector $ \bar{\mathbf{v}}_{s} $, given by
\begin{align}
	\bar{\mathbf{v}}_{s} = f_{\rm en}(\mathbf{G}^{\rm d}_{s})\in \mathbb{C}^{N_{\rm bit}\times 1},
\end{align}
where $ f_{\rm en} $ denotes the compression and quantization function. At the BS, the CSI is reconstructed from the limited feedback:
\begin{align}
	\hat{\mathbf{G}}^{\rm fb}_{s}=f_{\rm de}(\bar{\mathbf{v}}_{s})\in \mathbb{C}^{N_\text{RB}M_{\rm c}\times N_{\rm Rx}},
\end{align}
where $ f_{\rm de} $ denotes the corresponding dequantization and reconstruction function.
\subsubsection{CSI Acquisition Based on SRS}
SRS-based CSI acquisition depends on UL pilots and BS-side estimation, followed by DL mapping through reciprocity. As illustrated in the  \figref{systema}, SRS is necessarily configured in the FHS mode to amplify the received power compared to the full-band sounding \cite{zhu2024joint}.
In the FHS mode, the transmission bandwidth is divided into $L$ subbands, and each UE transmits pilots over only one subband during a given OFDM symbol. 
To support simultaneous CSI acquisition for multiple UEs, the comb-type transmission pattern is also adopted, where each UE transmits over $ N = N_{\rm c}/(LN_{\rm tc}) $ subcarriers within one OFDM symbol, and $N_{\rm tc}$ denotes the number of transmission combs.
It is assumed that each UE sounds the fullband channel $ K $ times through $ KL $ transmissions, then the received signal of SRS in the $n$-th subcarrier of $ l $-th subband at $k$-th  fullband sounding can be expressed by 
\begin{align}
	\mathbf{Y}^{\rm u}_{k,n,l} = \mathbf{H}^{\rm u}_{k,n,l} \mathbf{W}^{\rm u}\mathbf{X}^{\rm u}+\mathbf{N}^{\rm u}_{k,n,l}\in \mathbb{C}^{N_{\rm Tx}\times M_{\rm s}},\label{ultraining1}
\end{align}
where $\mathbf{H}^{\rm u}_{k,n,l} = \mathbf{H}\{S(kL+l),lN+n\} \in \mathbb{C}^{N_{\rm Tx}\times N_{\rm Rx}}$ denotes
the spatial domain channel in the $n$-th subcarrier of $ l $-th subband at $k$-th  fullband sounding, $ \mathbf{W}^{\rm u}\in \mathbb{C}^{ N_{\rm Rx}\times M_{\rm s}} $, $\mathbf{X}^{\rm u}\in \mathbb{C}^{M_{\rm s}\times M_{\rm s}} $, and $ \mathbf{N}^{\rm u}_{k,n,l}\in \mathbb{C}^{N_{\rm Tx}\times M_{\rm s}} $ denote the SRS pilot signal, the UE antenna pilot selection matrix, and noise, respectively. Similarly, the BS can estimate the low-dimensional equivalent channels as $ \mathbf{G}^{\rm u}_{k,n,l} = \mathbf{H}^{\rm u}_{k,n,l}\mathbf{W}^{\rm u} \in \mathbb{C}^{N_{\rm Tx}\times M_{\rm s}} $:
\begin{align}
	\hat{\mathbf{G}}^{\rm u}_{k,n,l} = f_{\rm ce}(\mathbf{Y}^{\rm u}_{k,n,l}),
\end{align}
where the $f_{\rm ce}$ denotes the channel estimation algorithm, which may be implemented using LS algorithm \cite{li2000optimum} or NN-based algorithms \cite{dong2019deep,zhou2023pay,he2024angle}. 
After that, the DL CSI can be inferred from estimated UL channels based on the channel reciprocity. 

In summary, CSI acquisition based solely on CSI-RS and SRS has been widely adopted in existing systems and continues to serve as the baseline solution for FR3 deployments. However, the unique characteristics of FR3 systems pose significant challenges to the effectiveness of using CSI-RS or SRS alone, which motivates further exploration of a joint UL and DL CSI acquisition scheme.
	\section{Joint UL and DL CSI Acquisition Framework \\
    under Practical System Configuration} \label{Section 3}
\begin{figure}[t]
		\centering
        \includegraphics[width=\linewidth,keepaspectratio]{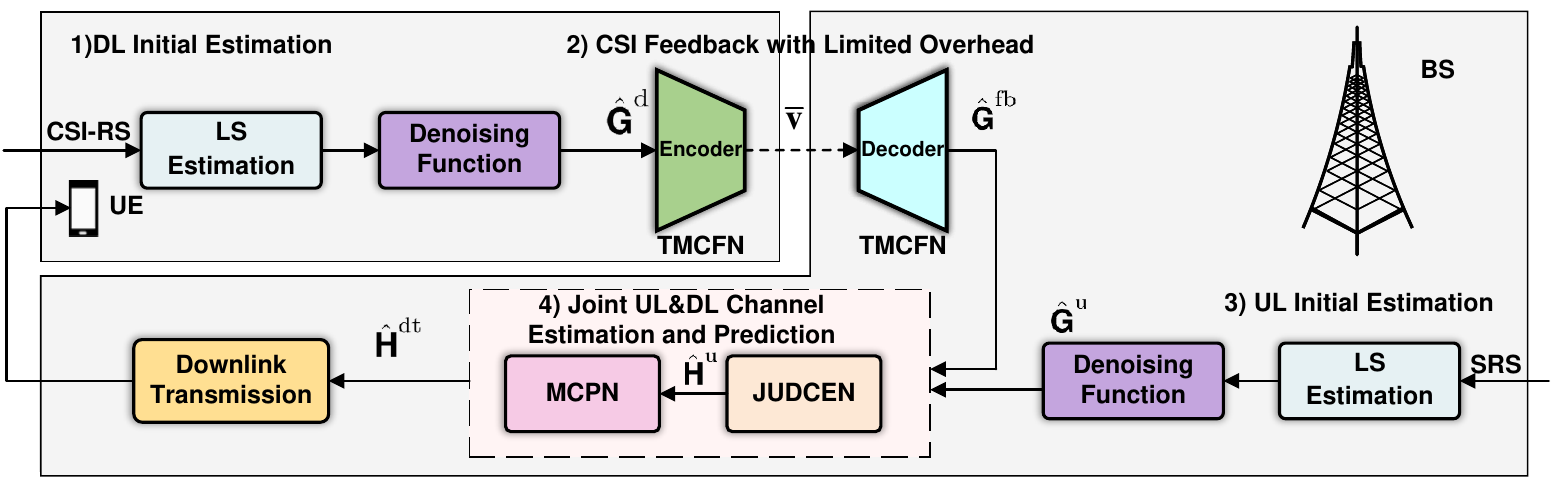}
		\caption{Joint UL and DL transmission framework.}
		\label{joint_csi_ce}
        \vspace{-5mm}
	\end{figure}

    To address the emerging challenges of FR3 transmissions, we propose a joint UL and DL CSI acquisition framework illustrated in Fig. \ref{joint_csi_ce}.
    In this framework, initial channel estimations are performed for both UL and DL links, where the estimated DL features are compressed, quantized, and fed back to the BS using the AI-based TMCFN module. Subsequently, the collected UL and DL channel features are jointly processed by the JUDCEN to reconstruct the full spatial-frequency domain CSI, which is followed by MCPN for channel prediction. By utilizing complementary observation structure and cross-referencing characteristics of CSI-RSs and SRSs, the proposed framework enables full-dimensional CSI reconstruction with low computational complexity and enhanced prediction performance across wide bandwidths.
    
\subsection{DL Initial Estimation} 
 Based on the received signal \eqref{dltraining}, we first perform the LS estimation of the DL channel over each slot and RB, which can be expressed as
	\begin{align}
    	\begin{aligned}
		\hat{\mathbf{G}}^{\rm d,LS}_{s,r}&=(\mathbf{X}^{\rm d})^{-1}\mathbf{Y}^{\rm d}_{s,r}\in  \mathbb{C}^{ M_{\rm c}\times N_{\rm Rx}}.
        	\end{aligned}
	\end{align}
To enhance the performance  under low signal-to-noise ratio (SNR) conditions, we further apply joint denoising (DN) estimation in the angle-delay domain by concatenating the LS estimation over all RBs. Specifically, the DL channel feature can be expressed by
	\begin{align}
        	\begin{aligned}
		\hat{\mathbf{G}}^{\rm d}_{s}=\mathbf{F}_{N_{\rm RB}}\otimes \mathbf{F}_{ M_{\rm c}}f_{\rm dn}(\mathbf{F}_{N_{\rm RB}}^{H}\otimes \mathbf{F}_{ M_{\rm c}}^{H}\hat{\mathbf{G}}^{\rm d,LS}_{s};\sigma^2_{\rm dl}),
                	\end{aligned}
	\end{align}
	where $\hat{\mathbf{G}}^{\rm d,LS}_{s}=[(\hat{\mathbf{G}}^{\rm d,LS}_{s,1})^T,(\hat{\mathbf{G}}^{\rm d,LS}_{s,2})^T,\ldots,(\hat{\mathbf{G}}^{\rm d,LS}_{s,N_{\rm RB}})^T]^T\in  \mathbb{C}^{N_{\rm RB}M_{\rm c}\times N_{\rm Rx}}$, $ f_{\rm dn} $ is the componentwise denoising function determined by the DL noise variance $ \sigma^2_{\rm dl} $, given by
    	\begin{align}
		[f_{\rm dn}\left(\mathbf{X};\sigma^2\right)]_{i,j}=[\mathbf{X}]_{i,j}\mu([\mathbf{X}]_{i,j}-\kappa\sigma),
	\end{align}
    where $\kappa$ and $\sigma^2$ are experimental parameter and noise variance, respectively, and $\mu(\cdot)$ denotes the Heaviside function. 
    
    By stacking the  DN-refined   DL channel feature $\hat{\mathbf{G}}^{\rm d}_{s}$ over $ N_{\rm dt} $ slots allocated for CSI-RS, the DL channel feature tensor $ {\hat{\tG}^{\rm d}}\in \mathbb{C}^{N_{\rm dt}\times N_{\rm RB} \times M_{\rm c}\times N_{\rm Rx}} $ can be constructed.  Compared to directly feeding the noisy DL received signal, the DN operation alleviates the representational burden on the subsequent feedback network, thereby facilitating more accurate decompression of the DL CSI.    	
    
\subsection{CSI Feedback with Limited Overhead} 
Following the initial DL channel estimation, CSI feedback is performed through a pipeline consisting of compression, quantization, transmission to the BS, and finally decompression.
Specifically, we first perform DFT along  spatial-frequency domain of DL channel features to exploit the inherent  sparsity, which can be expressed as \footnote{As the CSI feedback is performed independently in each slot, the slot index $s$ is omitted throughout this subsection for brevity.}
     \begin{align}
        \mathbf{H}^{\rm{ad}}=f_{\rm re}\left(\mathbf{F}_{N_{\rm RB}}\otimes \mathbf{F}_{ M_{\rm c}}\hat{\mathbf{G}}^{\rm d}_{}\mathbf{F}_{N_{\rm Rx}}\right)\in \mathbb{C}^{N_{\rm d} \times N_{\rm a}},\label{dft}
    \end{align}
    where $N_{\rm d}=N_{\rm RB}$ is the number of delay taps, $N_{\rm a}=M_{\rm c}N_{\rm Rx} $ is the  number of angle taps, and $f_{\rm re} $ denotes reshape operation. Building on this, an NN-based encoder is designed to compress the angle-delay domain channel, expressed as
	\begin{equation}
		\mathbf{v}= f_{\rm{e}}\left(\mathbf{H}^{\rm{ad}} ; \Theta_{\rm{e}}\right)\in \mathbb{R}^{M\times1},\label{fen}
	\end{equation}
	where $\Theta_{\rm{e}}$ denotes the parameters of the NN-based encoder. To comply with the quantized limited  feedback constraints imposed by current specifications, $ \mathbf{v}=[v_1,v_2,\ldots,v_M]^T $ is uniformly quantized to form a transmitting bits  $\bar {\bf v}$:
		\begin{align}
        		\begin{aligned}
&{\bar {\bf v}}_m=\chi\left(\rho\left\lfloor v_m/{\rho}\right\rfloor\right)\in \mathbb{R}^{\frac{N_{\rm bit}}{M}\times1}, \ m=1,...,M,\\
&\qquad\quad{\bar {\bf v}} = [{\bar {\bf v}}_1^T,{\bar {\bf v}}_2^T,\ldots,{\bar {\bf v}}_M^T]^T\in \mathbb{R}^{N_{\rm bit}\times 1},
	\end{aligned} \label{quan}
	\end{align}
	where  $\chi(\cdot)$ represents  bit conversion, $\lfloor\cdot\rfloor$ represents the rounding operation, $\rho=Z/2^{-N_{\rm bit}/{M}}$ is the quantization interval, $N_{\rm bit}$ is the number of quantization bits, and $Z =\max_{m} v_{m} - \min_{m} v_{m}$. We assume that the quantized bits are transmitted over a dedicated feedback link, allowing the BS to receive the feedback bits ${\bar {\bf v}}$ without loss \cite{guo2022overview}, which are then subjected to a dequantization process as follows
		\begin{align}
        		\begin{aligned}
&{\hat v}_m=\chi^{-1}\left({\bar {\bf v}}_m\right), \ m=1,...,M,\\
&\ \hat{\mathbf{v}}= [{\hat v}_1,{\hat v}_2,\ldots,{\hat v}_M]^T\in \mathbb{R}^{M\times 1}.
	\end{aligned}\label{dequan}
	\end{align}
    The resulting codeword $\hat{\mathbf{v}}$ is then decompressed using a NN-based decoder, and the final DL channel features is obtained by inverse DFT (IDFT):
	\begin{equation}	
\begin{aligned}
    		&\qquad\quad\qquad\quad{\hat{\mathbf{H}}^{\rm ad}}= f_{\rm{d}}\left(\hat{\mathbf{v}}
		; \Theta_{\rm{d}}\right)\in \mathbb{C}^{N_{\rm d} \times N_{\rm a}},\\ 
       &\hat{\mathbf{G}}^{\rm fb}=f_{\rm re}\left(\mathbf{F}_{N_{\rm RB}}^{H}\otimes \mathbf{F}_{ M_{\rm c}}^{H}{\hat{\mathbf{H}}^{\rm ad}}\mathbf{F}_{N_{\rm Rx}}^H\right)\in  \mathbb{C}^{M_{\rm c}N_{\rm RB}\times N_{\rm Rx}},
\end{aligned}
        \label{fde}
	\end{equation}
	where  $\Theta_{\rm{d}}$ denotes the parameters of the NN-based decoder. By combining \eqref{dft}–\eqref{fde} sequentially, the CSI feedback process can be formulated as follows
\begin{align}
		{\hat{\tG}^{\rm fb}}=f_{\rm TMCFN}({\hat{\tG}^{\rm d}})\in \mathbb{C}^{N_{\rm dt}\times N_{\rm RB} \times M_{\rm c}\times N_{\rm Rx}},\label{TMCFN}
	\end{align}
	where ${\hat{\tG}^{\rm fb}}$ denotes the DL channel feature tensor obtained at the BS through CSI feedback and $f_{\rm TMCFN}$ is the function implemented by  TMCFN.
    \subsection{UL Initial Estimation}
 Similarly, LS-based estimation of the UL SRS is performed for each slot and subcarrier, followed by the DN-based refinement in the angle-delay domain, expressed as
    	\begin{align}
        	\begin{aligned}
         &   \quad	\qquad 	\hat{\mathbf{G}}^{\rm u,LS}_{k,n,l}= \mathbf{Y}^{\rm u}_{k,n,l}(\mathbf{X}^{\rm u})^{-1}\in  \mathbb{C}^{ N_{\rm Tx}\times M_{\rm s}},\\
		&		\hat{\mathbf{G}}^{\rm u}_{k,l}=\mathbf{F}_{N}\otimes \mathbf{F}_{N_{\rm Tx}}f_{\rm dn}\left(\mathbf{F}_{N}^{H}\otimes \mathbf{F}_{N_{\rm Tx}}^{H}\hat{\mathbf{G}}^{\rm u,LS}_{k,l};\sigma^2_{\rm ul}\right),
                	\end{aligned}
	\end{align}
	where $\hat{\mathbf{G}}^{\rm u,LS}_{k,l}=[(\hat{\mathbf{G}}^{\rm u,LS}_{k,1,l})^T,\ldots,(\hat{\mathbf{G}}^{\rm u,LS}_{k,N,l})^T]^T\in  \mathbb{C}^{NN_{\rm Tx}\times M_{\rm s}}$, and $ \sigma^2_{\rm ul} $ denotes the UL noise variance. By stacking the DN-refined  UL channel features $\hat{\mathbf{G}}^{\rm u}_{k,l}$ across all  $ L $ subbands and $ K $ full-band sounding instances, the UL channel feature tensor $ \hat{\tG}^{\rm u}\in \mathbb{C}^{K\times L \times N \times N_{\rm Tx}\times M_{\rm s}} $ can be constructed. The DN operation for the UL initial estimation also reduces the representational burden on the subsequent feedback network and facilitates more accurate decompression of the UL CSI.

\subsection{Joint UL and DL Channel Estimation and Prediction}	

In FR3 systems, the number of antennas at the BS far exceeds the number of transmit ports supported by current specifications \cite{3gpp.38.211}. As a result, both DL and UL initial estimations can capture only partial observations of the spatial-domain channel, making it difficult to reconstruct accurate CSI when either direction is used alone. This limitation motivates the joint exploitation of UL and DL signals to achieve more complete CSI acquisition. 

\begin{figure}[tp!] 
		\centering  
		\subfigtopskip=2pt
		\subfigbottomskip=2pt 
		\subfigcapskip=-5pt 
		\subfigure[]{
		 	\includegraphics[width=0.45\linewidth]{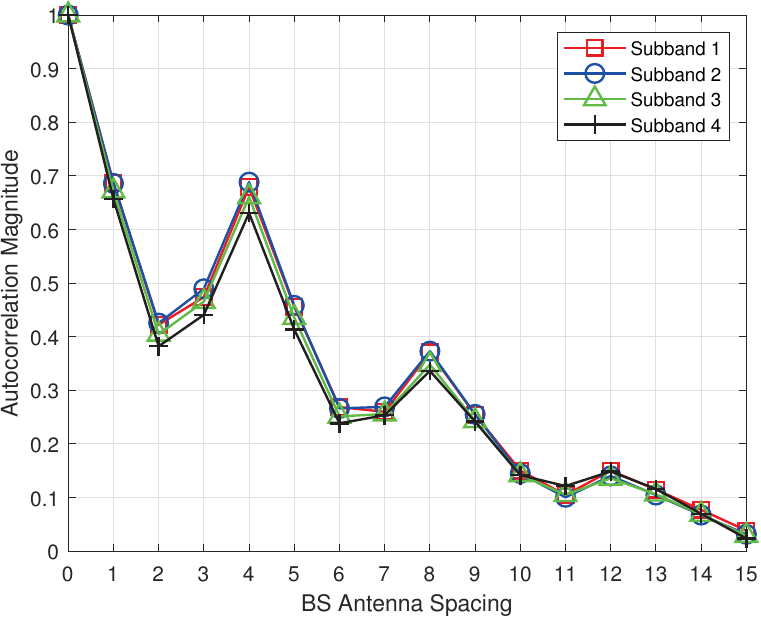}\label{BSAnt_cor}}
            \subfigure[]{		 		\includegraphics[width=0.45\linewidth]{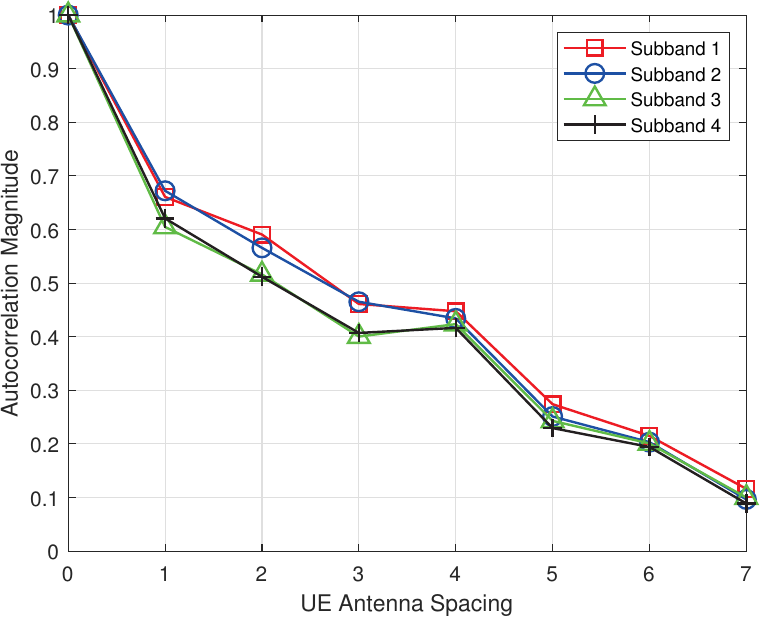}\label{UEAnt_cor}}
		 \caption{Autocorrelation function of spatial domain channel across:(a) BS antennas; (b) UE antennas.}
        \label{UEBSAnt_cor}
    \end{figure}

Fortunately, a complementary spatial probing structure inherently exists between DL and UL channels: the DL probing provides full spatial CSI at the UE side but only partial observations at the BS side, while UL probing exhibits the complementarity pattern, as shown in \figref{diff_csirs_srs_v1}. 
Motivated by this complementary spatial information, we further examine the spatial characteristics of the channels. Specifically, the spatial autocorrelation functions of the BS and UE are expressed as $R_{\rm BS}(x)=\operatorname{E}\left((\mathbf{h}^{H}_{(i+x):}\mathbf{h}_{i:})/({\|\mathbf{h}_{(i+x):}\|_2\|\mathbf{h}_{i:}\|_2})\right), R_{\rm UE}(x)=\operatorname{E}\left((\mathbf{h}^{H}_{:(j+x)}\mathbf{h}_{:j})/({\|\mathbf{h}_{:(j+x)}\|_2\|\mathbf{h}_{:j}\|_2})\right)$, where $\mathbf{h}_{i:}$ and $\mathbf{h}_{:j}$ denote the $i$-th row and $j$-th column of $\mathbf{H}\{n,k\}$ defined in \eqref{H_d}, respectively.
As shown in \figref{BSAnt_cor} and \figref{UEAnt_cor}, it reveals strong spatial correlation at both ends. These observations statistically validate the feasibility of leveraging CSI-RS and SRS measurements jointly to mitigate the partial spatial probing problem.
Based on this insight, we propose JUDCEN to fuse UL and DL channel features:
\begin{align}
	\hat{\tH}^{\rm u}=f_{\rm JUDCEN}({\hat{\tG}^{\rm fb}},{\hat{\tG}^{\rm u}})\in \mathbb{C}^{K\times L \times N \times N_{\rm Tx}\times N_{\rm Rx}},
\end{align}
where $ \hat{\tH}^{\rm u} $ and $ f_{\rm JUDCEN} $ denote the recovered UL channels and function implemented by JUDCEN, respectively.
Within this network, a dedicated shape-matching module is introduced to address the frequency domain mismatch in probing patterns, as illustrated in Fig.~\ref{diff_csirs_srs_v1}, and detailed in Section~\ref{Section 4}. This module specifically reshapes the observed DL and UL channel features subband-wise along the frequency domain, enabling the downstream fusion module to operate on aligned representations. It is important to note that unobserved subbands are inferred by the subsequent channel prediction network. To mitigate the information loss caused by limited CSI feedback, we also propose a deep learning-assisted CSI feedback scheme, termed TMCFN, which integrates into the overall end-to-end joint optimization framework. Notably, the proposed framework exploits the inherent CSI-RS signaling in TDD systems \cite{3gpp.38.211} without introducing additional pilot overhead.
         
 After the UL-DL CSI fusion, the rapid temporal channel variations in FR3 necessitate explicit channel prediction schemes to ensure reliable CSI acquisition. To this end, the proposed MCPN  models long-range dependencies across multiple subbands to predict intermediate channels between two consecutive CSI acquisitions. Specifically, the DL transmission channels over $ N_{\rm tr} $ slots are predicted by
\begin{align}
	\hat{\tH}^{\rm dt}=f_{\rm MCPN}(\hat{\tH}^{\rm u})\in \mathbb{C}^{N_{\rm tr}\times L \times N \times N_{\rm Tx}\times N_{\rm Rx}},
\end{align}
where $f_{\rm MCPN}$ is the function implemented by MCPN. The details of the proposed networks, including TMCFN, JUDCEN, and MCPN, will be presented in the following section.
\newpage
\section{Hybrid Attention- and SSM-Based \\CSI Acquisition Network Design} \label{Section 4}
In the proposed joint UL and DL CSI acquisition framework, the TMCFN, JUDCEN, and MCPN  are designed to achieve CSI feedback, joint UL and DL channel estimation, and channel prediction, respectively. This section presents the detailed network architectures that collectively constitute the proposed HASCAN framework:
\begin{itemize}
\item \textbf{TMCFN}: It implements sparse-domain CSI feedback by integrating Transformer and MLP architectures, which exploits the angle-delay domain sparsity to achieve efficient and accurate compression and reconstruction.
\item \textbf{JUDCEN}: It fuses UL and DL channel features for spatial-frequency domain reconstruction via a spatial-frequency attention mechanism, which effectively enhances feature interaction between UL and DL channels.
\item \textbf{MCPN}: It is responsible for multi-subband channel prediction in the angle-delay domain using stacked Mamba blocks, enabling effective modeling of long-range temporal dependencies with low computational complexity.
\end{itemize}
In what follows, we provide an in-depth description of the proposed networks, highlighting their architectural designs, key modules, and respective roles in the joint UL and DL CSI acquisition framework.
\subsection{Transformer-MLP-Based CSI Feedback Network}
	\begin{figure}[t]
		\centering
		\includegraphics[width=3.53in]{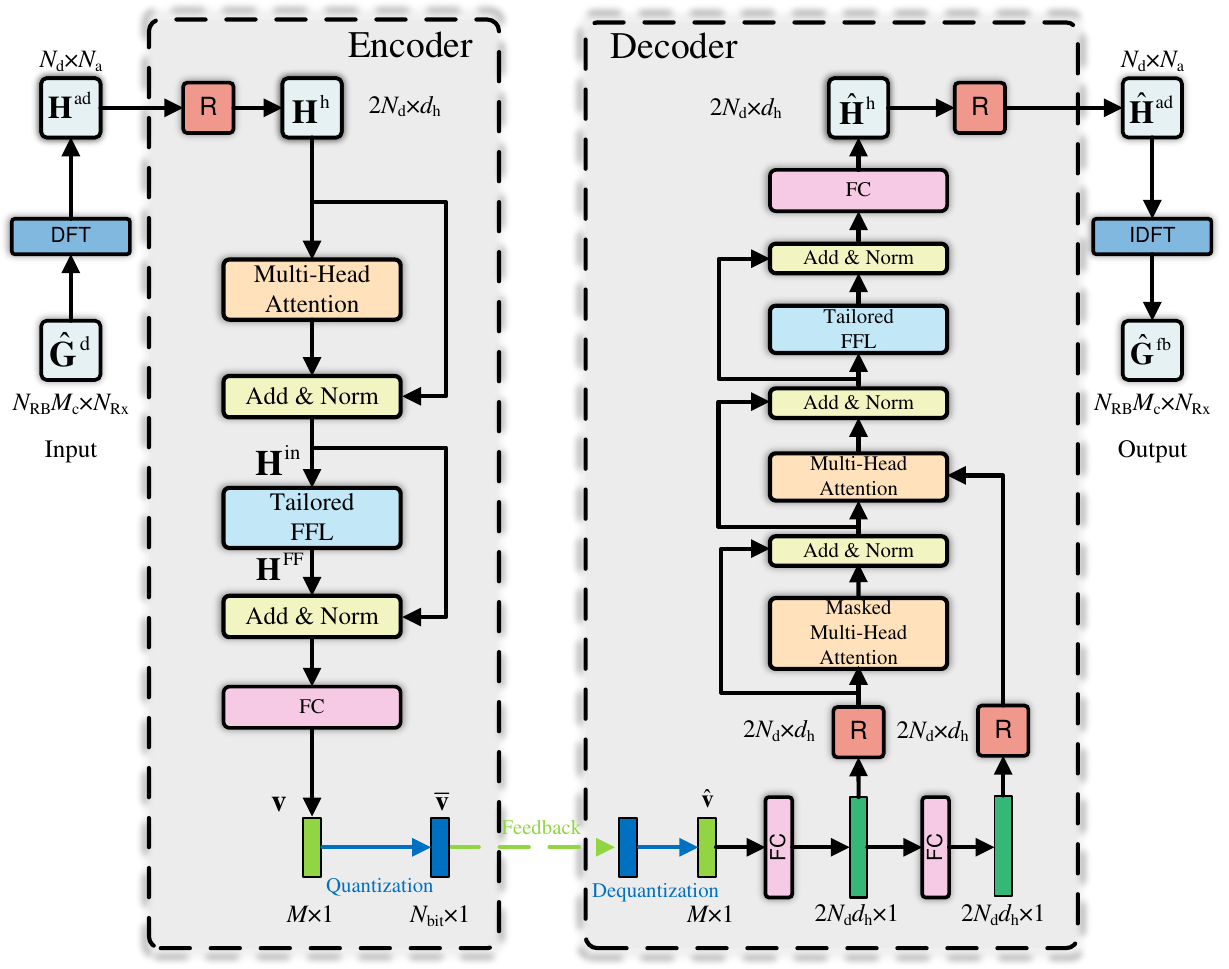} 
		\caption{Transformer-MLP-based CSI feedback network.}
		\label{CovNet_0911} 
        \vspace{-5mm}
	\end{figure}
    With multi-head self-attention mechanism, the Transformer architecture \cite{vaswani2017attention} can effectively model global sequence dependencies and is thus well suited for sequence prediction. Therefore, we adopt this architecture to design the CSI feedback network as its strength in modeling sequence dependencies naturally  aligns with our treatment of the angle-delay domain channel as a set of multiple delay-domain sequences. Furthermore, since the FFL in the standard Transformer can only perform feature transformation along a single dimension, we upgrade it with an MLP-inspired design \cite{tolstikhin2021mlp} by adopting the partial architecture in \cite{zhuang2024covnet},  which enables joint exploitation of feature interactions in the angle–delay domain.
    
	Specifically, the TMCFN consists of a multi-head attention-based encoder at the UE side and a masked multi-head attention-based decoder at the BS side, as shown in \figref{CovNet_0911}. The angle-delay domain channel is initially processed by the attention layer. Denote the input of the attention layer as $ \mathbf{H}^{\rm h} \in \mathbb{R}^{2N_{\rm d} \times d_{\rm h}} $, where the 2 represents the complex dimension,  and the representation dimension is given by  $d_{\rm h}=M_{\rm c}N_{\rm Rx} $, the self-attention matrix $\mathbf{A}_i \in \mathbb{R}^{2N_{\rm d} \times d_v} $  can be expressed by
	\begin{align}
		\begin{aligned}
            &\mathbf{A}_i=\operatorname{Attention}\left(\mathbf{H}^{\rm h} \mathbf{W}_i^Q, \mathbf{H}^{\rm h}\mathbf{W}_i^K,\mathbf{H}^{\rm h} \mathbf{W}_i^V\right),\\
			&\ \operatorname{Attention}\left(\mathbf{Q}, \mathbf{K}, \mathbf{V}\right)=\operatorname{softmax}\left(\frac{\mathbf{Q K}^T}{\sqrt{d_k}}\right) \mathbf{V}, \label{ohead}
		\end{aligned}
	\end{align}
    where the $\mathbf{W}_i^Q \in \mathbb{R}^{d_{\rm h} \times d_k}, \mathbf{W}_i^K \in \mathbb{R}^{d_{\rm h} \times d_k}$, and $\mathbf{W}_i^V \in \mathbb{R}^{d_{\rm h} \times d_v}$ denote query, key, and value learnable matrices associated with the $i$-th attention head, $i=1,2,\ldots N_{\rm m}$, respectively, projecting the input  $ \mathbf{H}^{\rm h}  $ to  different representation subspaces,  $ d_k = d_v = d_{\rm h}/N_{\rm m}  $, and $\operatorname{softmax}\left(\cdot\right)  $ denotes the softmax function. The attention layer employs a multi-head self-attention mechanism \cite{vaswani2017attention} to capture  inter-delay dependencies across multiple propagation paths. The outputs from all attention heads are concatenated and linearly transformed as
    \begin{align}
    {\bar {\mathbf{H}}^{\rm h}}=\left[\mathbf{A}_1, \ldots, \mathbf{A}_{N_{\rm m}}\right] \mathbf{W}^O\in \mathbb{R}^{2N_{\rm d} \times d_{\rm h}},
    \end{align}
	where $ \mathbf{W}^O\in \mathbb{R}^{d_{\rm h} \times d_{\rm h}} $ is also a learnable matrix.

Subsequently, the attention layer output is processed through a residual connection (RN) followed by the layer normalization (LN).  To further enhance feature interaction modeling in the angle-delay domain, we upgrade the FFL in the Transformer architecture to the tailored feed-forward layer (TFFL). Specifically, the input $\mathbf{H}^{\rm in}\in \mathbb{R}^{2N_{\rm d} \times d_{\rm h}} $ to the TFFL is first expanded into a square matrix via an FC layer, which can be formulated by
	\begin{align}
		\mathbf{H}^{\rm S}=\mathbf{H}^{\rm in} \mathbf{W}^{G 1}+(\mathbf{b}^{G 1})^T\in \mathbb{R}^{2N_{\rm d} \times 2N_{\rm d}},
	\end{align}
	where $\mathbf{W}^{G 1} \in \mathbb{R}^{d_{\rm h} \times 2N_{\rm d}}$ and $\mathbf{b}^{G 1} \in \mathbb{R}^{2N_{\rm d}\times 1}$ are learnable weight matrices and bias vectors, respectively. Then, the expanded representation  $\mathbf{H}^{\mathrm{SC}}\in \mathbb{R}^{2N_{\rm d} \times d_{\rm f}}$ is obtained by passing both the square matrix $\mathbf{H}^{\mathrm{S}}$ and its transpose followed by concatenation, which is then utilized to produce the final output of the TFFL layer:
    	\begin{align}
		\begin{aligned}
\mathbf{H}^{\mathrm{SC}}&=[(\mathbf{H}^{\mathrm{S}} \mathbf{W}^{G 2}+(\mathbf{b}^{G 2})^T) ,((\mathbf{H}^{\mathrm{S}})^T \mathbf{W}^{G 3}+(\mathbf{b}^{G 3})^T)],\\
            			\mathbf{H}^{\mathrm{FF}}&= \operatorname{Drop}(\operatorname { R e L U } (\mathbf{H}^{\mathrm{SC}});p) \mathbf{W}^{G 4}+(\mathbf{b}^{G 4})^T\in \mathbb{R}^{2N_{\rm d} \times d_{\rm h}},
		\end{aligned}
	\end{align}
	where $\operatorname{ReLU}(\cdot)$ and $\operatorname{Drop}(\cdot;p)$ denote the rectified linear unit (ReLU) activation function and dropout operation, respectively, $p$ denotes the dropout probability,    $\mathbf{W}^{G 2},\mathbf{W}^{G 3} \in \mathbb{R}^{2N_{\rm d} \times d_{\rm f}/2}$ and $\mathbf{W}^{G 4} \in \mathbb{R}^{d_{\rm f} \times d_{\rm h}}$ are learnable weight matrices, $d_{\rm f}$ is the inner dimension of the TFFL,  $\mathbf{b}^{G 2}, \mathbf{b}^{G 3} \in \mathbb{R}^{d_{\rm f}/2\times 1}$ and $\mathbf{b}^{G 4} \in \mathbb{R}^{d_{\rm h}\times 1}$ are learnable bias vectors. Then, the output of the TFFL $\mathbf{H}^{\mathrm{FF}}$ is  processed through another RN followed by LN.
    
    Subsequently, the codeword $\mathbf{v}$ is obtained from the final FC layer of the encoder, followed by quantization, feedback, and dequantization \eqref{quan}-\eqref{dequan}, which is then served as the input to the decoder. Similar to the encoder structure, the output of the multi-head attention module is fed into an FC layer with a reshape operation to obtain the estimated ${\hat{\mathbf{H}}^{\rm ad}}$.
    
\subsection{Spatial-Frequency-Attention-Based Channel Estimation Network} \label{SAM} 
With the UL and DL features obtained from UL and DL estimation and  feedback, the BS then employs JUDCEN, which consists of a preprocessing module and a spatial-frequency-attention-based module, to combine them for full CSI reconstruction. To leverage complementary observation and cross-referencing characteristic of CSI-RSs and SRSs in the spatial–frequency domain as mentioned before, the preprocessing module is designed to first fuse UL and DL features in the spatial domain and project them into a feature tensor suitable for final channel estimation. Moreover, the spatial-frequency-attention mechanism, guided by attention maps produced through pooling and convolution, adaptively enhances critical regions in the feature map and suppresses redundant background responses, which is suitable for image recognition, detection, and segmentation. Considering that the spatial-domain channel exhibits 2D features similar to images, we employ spatial-frequency attention \cite{woo2018cbam} to perform pooling and convolution along the frequency domain, thereby generating the two-dimensional spatial attention maps for each channel sample. This guides the network to achieve precise estimation of the spatial-frequency domain channel.

Initially, we construct the input of the JUDCEN, since the UL and DL features hold different distributions in frequency domain as exhibited in \figref{systemb}. Specifically, we  reshape the CSI-RS signals $ {\hat{\tG}^{\rm fb}} $ to match the dimensionality of the SRS signals $ \hat{\tG}^{\rm u} $ for subsequent tensor operations, given by
	\begin{align}
		{\hat{\tG}^{\rm s}}=f_{\rm sm}({\hat{\tG}^{\rm fb}})\in  \mathbb{C}^{K\times L \times N \times N_{\rm Tx}\times N_{\rm Rx}},
	\end{align}
	where $ f_{\rm sm} $ denotes the shape matching function. The shape matching function duplicates each CSI-RS signal within an RB $ M_{\rm sc}/N_{\rm tc} $ times to align with the frequency-domain resolution of the SRS.  Subsequently, the shape-matched CSI-RS and SRS signals are  concatenated along the spatial dimension, forming the input tensor $ \hat{\tG}^{\rm ud} \in \mathbb{R}^{K\times L \times N \times 2N_{\rm Tx} ( N_{\rm Rx}+  M_{\rm s})} $, where the 2 represents the complex dimension. 
    
\begin{figure*}
		\centering
		\includegraphics[width=7in]{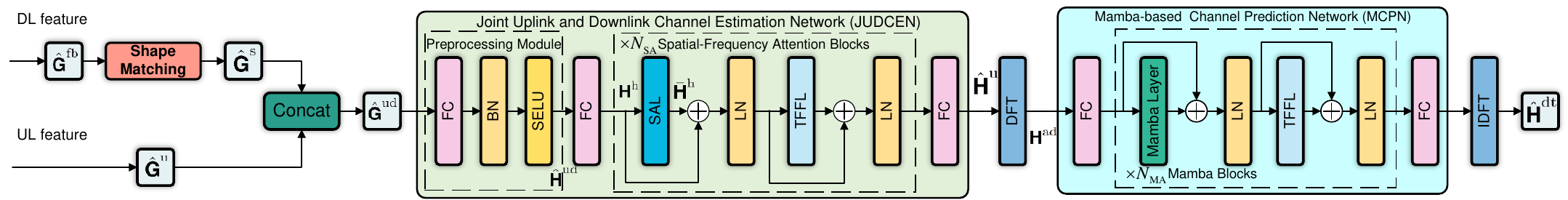}
		\caption{{The overall architecture of the JUDCEN and MCPN.}
		}
          \vspace{-5mm}
		\label{Network}
	\end{figure*}

    The concatenated feature $ \hat{\tG}^{\rm ud}$ is first addressed by the preprocessing module, composed of FC layers, batch normalization (BN), and SELU activations, which is projected into a target space suitable for subsequent processing, given by
	\begin{align}
		\hat{\tH}^{\rm ud}=\operatorname{SELU}(\operatorname{BN}(\operatorname{FC}_{2N_{\rm Tx} N_{\rm Rx}}(\hat{\tG}^{\rm ud})))\in \mathbb{R}^{ K\times L \times N \times 2N_{\rm Tx} N_{\rm Rx} },
	\end{align}
    where the  $\operatorname{FC}_{x}$ is a parameterized linear projection to dimension $x$. Then, the spatial-frequency attention module is mainly composed of $ N_{\rm SA} $ spatial-frequency attention blocks (SABs), and the additional FC layers are used for dimensionality augmentation and reduction. As shown in Fig. \ref{Network}, each SAB consists of one spatial-frequency attention layer (SAL), one FFL, and two LN layers with several RNs. Denote the input of the attention layer as $ \tH^{\rm h}\in \mathbb{R}^{ d_{\rm f} \times N_{\rm Rx} \times N_{\rm Tx} }$, the SAL applies both average-pooling and max-pooling along the frequency domain to obtain two initial feature maps of spatial channel:
         \begin{align}
    \tP= \operatorname{Concat}\left(\operatorname{AvgP}\left(\tH^{\rm h}\right); \operatorname{MaxP}  \left(\tH^{\rm h}\right)\right)\in \mathbb{R}^{ 2\times N_{\rm Rx} \times N_{\rm Tx} },
    \end{align}
    where the $\operatorname{Concat}(\cdot)$ denotes the concatenating operation. Then, the final spatial feature map $\mathbf{P}$ is generated through one $ 1\times1\times2 $ convolutional filter followed by sigmoid activation function on the concatenated feature map $\tP$:
        	\begin{align}
		\begin{aligned}
\mathbf{P}=\sigma\left(\operatorname{Conv}\left(\tP\right)\right)\in \mathbb{R}^{ N_{\rm Rx} \times N_{\rm Tx} }.
		\end{aligned} \label{maps}
	\end{align}
     The spatial feature $\mathbf{P}$ is the essence of the SAL, which guides the network to focus on the dominant channel elements by:
        	\begin{align}
		\begin{aligned}
			   \bar{\tH}^{\rm h}=\tP^{[d_{\rm f}]} \odot \tH^{\rm h}\in \mathbb{R}^{ d_{\rm f} \times N_{\rm Rx} \times N_{\rm Tx} }.
		\end{aligned} \label{AAM}
	\end{align}
    where $\tP^{[d_{\rm f}]}\in \mathbb{R}^{ d_{\rm f} \times N_{\rm Rx} \times N_{\rm Tx} }$ is obtained by replicating $\mathbf{P}$ along the first dimension. In summary, by first performing UL and DL feature fusion in the spatial domain through the preprocessing module, the spatial-frequency-attention mechanism can then apply pooling and convolution along the frequency domain to generate spatial attention maps. This enables the network to conduct attention-guided channel estimation, facilitating subsequent channel prediction.
\subsection{Mamba-Based Channel Prediction Network} \label{Mamba}
 For prediction tasks, the self-attention mechanism and positional encoding in Transformer enable global information awareness and flexible capture of sequential dependencies, while multi-head attention significantly enhances feature representation. However, Transformer faces inherent limitations, particularly the quadratic computational complexity of attention mechanisms and challenges in handling long-range dependencies, leading to excessive complexity and degraded accuracy in multi-subband channel prediction \cite{dai2019transformer}. Considering that the sequence length involved in channel prediction is much longer than that in CSI feedback $2KNL\gg 2N_{\rm d}$, we employ the SSM-based Mamba layer as the primary block for constructing the prediction network. Specifically, Mamba achieves efficient sequence modeling through linear recurrence to maintain linear computational complexity, and it retains long-term historical state information through a sophisticated state transition matrix, thereby demonstrating strong long-range dependency modeling capability \cite{gu2024mamba}.

  \subsubsection{Architectures}
\begin{figure}[t]
    \centering
    \includegraphics[width=3.45in]{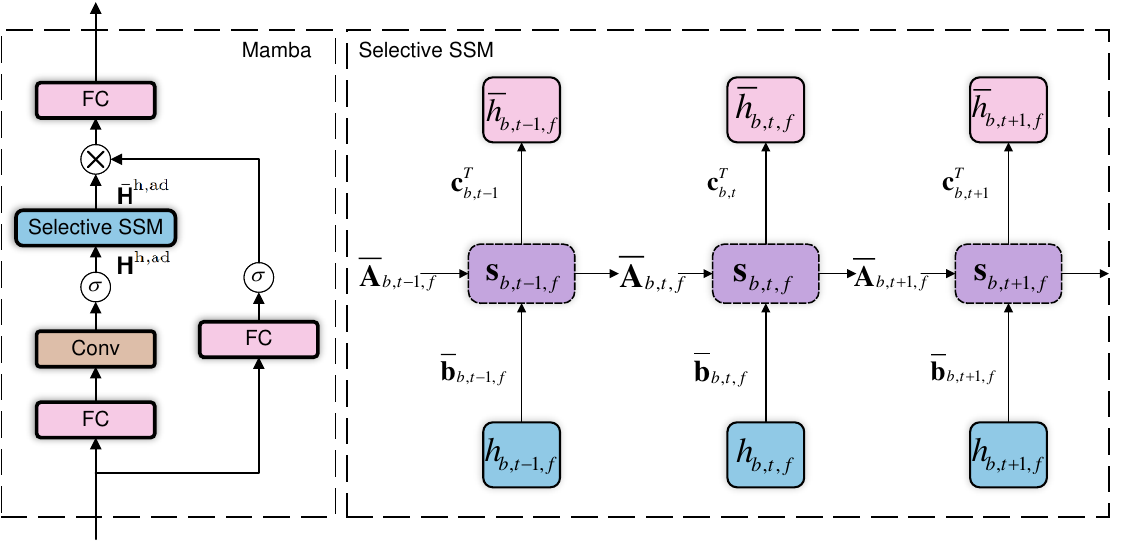}
    \caption{The structure of Mamba layer and selective SSM.}
    \label{ssmfigure}
    \vspace{-5mm}
\end{figure}

 The proposed MCPN structure is mainly composed of $ N_{MA} $ stacked Mamba blocks. Specifically, the basic structure of Mamba block is illustrated in \figref{Network}, which consists of one mamba layer, one FFL, and two LN layers with several residual links. Considering the advantage of the Mamba network in capturing long-sequence features, we first stack subcarriers from different subbands and apply DFTs in the frequency and spatial domains. This transformation enables subsequent multi-subband channel prediction via the MCPN in the angle-delay domain, and the resulting output is denoted as $ \tH^{\rm ad}\in \mathbb{R}^{ B \times N_{\rm sq} \times (N_{\rm Tx}N_{\rm Rx}) }$, where $B$ denotes the batch size, $N_{\rm sq}=2KNL$ represents sequence length, and $2$ denotes the complex dimension. The Mamba layer is built upon the H3 architecture \cite{fu2023hungry}, which serves as the foundation for many well-known SSM structures. By further incorporating the gated attention unit from Gated MLP \cite{hua2022transformer}, the two components are combined into a uniformly stacked module, as illustrated in Fig. \ref{ssmfigure}.  To enhance feature representation of the selective SSM module, the angle-delay domain channel tensor $ \tH^{\rm ad}$ is first mapped through an FC layer into $ \tH^{\rm h,ad}\in \mathbb{R}^{ B \times N_{\rm sq} \times d_{\rm f} }$, which is then processed by 
\begin{align}
    \bar{\tH}^{\rm h,ad}=\operatorname{SSSM}(\tH^{\rm h,ad};g(\tH^{\rm h,ad}))\in \mathbb{R}^{ B \times N_{\rm sq} \times d_{\rm f} },
    \label{sssm}
\end{align}
where $\operatorname{SSSM}$ denotes the operation of the selective SSM, $g(\tH^{\rm h,ad})=\{\bar{\tA},\bar{\tB},\tC\}$ denotes the parameter generation function, and $\bar{\tA}\in \mathbb{R}^{ B \times N_{\rm sq} \times d_{\rm f} \times N_{\rm H}}$, $\bar{\tB}\in \mathbb{R}^{ B \times N_{\rm sq} \times d_{\rm f} \times N_{\rm H}}$, $\tC\in \mathbb{R}^{ B \times N_{\rm sq}  \times N_{\rm H}}$ denote the generated parameter tensors that adapt to each input sample in the batch and time step, thereby enabling selective modeling. To clarify the internal computation of the selective SSM, we now detail its element-wise formulation, given by
\begin{align}
    \bar{h}_{b,t,f}=\operatorname{SSM}(\{h_{b,j,f}\}_{j=1}^{t};\{\bar{\mathbf{A}}_{b,j,f},\bar{\mathbf{b}}_{b,j,f},\mathbf{c}_{b,j}\}_{j=1}^{t}),
    \label{sssm_d}
\end{align}
where $\operatorname{SSM}(\cdot)$ denotes the discrete SSM with input-dependent parameter sets; $h_{b,t,f}$ and $\bar{h}_{b,t,f}$ denote the $(b,t,f)$-th elements of $\tH^{\rm h,ad}$ and $\bar{\tH}^{\rm h,ad}$, respectively; $\bar{\mathbf{b}}_{b,t,f} = \bar{\mathbf{b}}_{btf:} \in \mathbb{R}^{N_{\rm H} \times 1}$ and $\mathbf{c}_{b,t} =  \mathbf{c}_{bt:} \in \mathbb{R}^{N_{\rm H} \times 1}$ denote the $(b,t,f)$-th fiber vectors of $\bar{\tB}$ and $\tC$, respectively; $\bar{\mathbf{A}}_{b,t,f}=\operatorname{diag}\{\bar{\mathbf{a}}_{b,t,f}\}$ with   $\bar{\mathbf{a}}_{b,t,f}=\bar{\mathbf{a}}_{btf:} \in \mathbb{R}^{N_{\rm H} \times 1}$.
Subsequently, we introduce the design of the selective SSM module, which is developed from the conventional SSM through standard state-space formulation, discretization with convolution, and the incorporation of a selective mechanism.

    \subsubsection{State Space Model}
The structured state space sequence (S4) model \cite{gu2022efficiently}, inspired by continuous-time dynamical systems, maps an input signal $ h(t)\in \mathbb{R}$ to $\bar{h}(t)\in \mathbb{R} $ via an implicit latent state $ \mathbf{s}(t)$. It provides a principled framework for modeling physical processes, particularly linear time-invariant (LTI) systems, by leveraging a learnable transition matrix $ \mathbf{A}\in \mathbb{R}^{N_{\rm H}\times N_{\rm H}} $, along with learnable input and output projection vectors, $ \mathbf{b}\in \mathbb{R}^{N_{\rm H}\times 1} $ and $ \mathbf{c}\in \mathbb{R}^{N_{\rm H}\times 1} $. This model is described by the following state-space representation
	\begin{align}
		\begin{aligned}
			\frac{{\rm d}\mathbf{s}(t)}{{\rm d}t}&=  \mathbf{A}\mathbf{s}(t)+ \mathbf{b} h(t),\\
			\bar{h}(t)&=\mathbf{c}^T\mathbf{s}(t). \label{s4}
		\end{aligned}
	\end{align}
    
    To effectively model long-range dependencies, the structure of  $\mathbf{A}$ is of critical importance. Within this representation, Mamba adopts a diagonal formulation of $\mathbf{A} = \operatorname{diag}\{\mathbf{a}\}$, following the S4D-Real structure \cite{gu2022parameterization} and HIPPO theory \cite{gu2020hippo}. To enable historical information retention, the diagonal elements are initialized as $[\mathbf{a}]_i = -(i+1)$ for $i = 1, \ldots, N_{\rm H}$.
    
        \subsubsection{Discretization and Convolution}
   Instead of applying the S4 model directly to a continuous function $h(t)$, we discretize \eqref{s4} using a step size $d_f$, which defines the input resolution. Accordingly, the continuous parameters $\mathbf{A}_f$ and $\mathbf{b}_f$ are discretized as $\bar{\mathbf{A}}_f=f_A(d_f,\mathbf{A}_f), \bar{\mathbf{b}}_f=f_B(d_f,\mathbf{A}_f, \mathbf{b}_f)$, where $f_A$ and $f_B$ denote the discretization rules, given by the zero-order hold methods as follows
	\begin{align}
        \begin{aligned}
    		f_A(d_f,\mathbf{A}_f) &=  \operatorname{exp}(d_f\mathbf{A}_f), \label{dis} \\
    		f_B(d_f,\mathbf{A}_f, \mathbf{b}_f) &=(\mathbf{A}_f)^{-1}(\operatorname{exp}(d_f\mathbf{A}_f)-\mathbf{I}_{N_{\rm H}})\mathbf{b}_f, 
        \end{aligned}
	\end{align}
	After converting the continuous parameters $ (d_f,\mathbf{A}_f,\mathbf{b}_f) $ into their discrete equivalents $ (\bar{\mathbf{A}}_f,\bar{\mathbf{b}}_f) $ via \eqref{dis}, the output sequence $\bar{h}_{b,t,f}$ can be computed by discrete SSM
	\begin{align}
		\begin{aligned}
			\mathbf{s}_{b,t,f}&=  \bar{\mathbf{A}}_{f}\mathbf{s}_{b,t-1,f}+ \bar{\mathbf{b}}_{f} h_{b,t,f},\\
			\bar{h}_{b,t,f}&=\mathbf{c}_{f}^{T}\mathbf{s}_{b,t,f}, \label{rnn}
		\end{aligned}
	\end{align}
    which can be simplified into a convolutional summation:
	\begin{align}
		\begin{aligned}
\textstyle \bar{h}_{b,t,f}=\sum_{j=1}^{t}\mathbf{c}_{f}^T\bar{\mathbf{A}}^{t-j}_{f}\bar{\mathbf{b}}_{f}h_{b,j,f}. \label{cov}
		\end{aligned}
	\end{align}
     As shown above, the SSM-based network can be efficiently computed in either a recurrent or convolutional form, enabling linear or near-linear scaling with respect to sequence length. Compared to the Transformer architecture, the computational complexity is reduced from $\mathcal{O}(N_{\rm sq}^2)$ to $\mathcal{O}(N_{\rm sq})$ \cite{martin2018parallelizing}.
\subsubsection{Selective SSM}
 However, the S4 model has limited sequence modeling capacity due to the time-invariant and input-independent parameterization, i.e.,  these parameters  $\{\mathbf{A}_{f},\mathbf{b}_{f},\mathbf{c}_{f}\}_{f=1}^{d_{\rm f}}$ are independent of both $t$ and $b$.
 To overcome this limitation, Mamba introduces selective SSM to add time-variant and input-dependent dynamics. Specifically, the state-space tensors in selective SSM are conditioned on the input tensor $\tH^{\rm h,ad}\in \mathbb{R}^{B\times N_{\rm sq}\times d_{\rm f}}$, and are parameterized as
\begin{align}
    \begin{aligned}
    \tB&=s_{B}(\tH^{\rm h,ad})\in \mathbb{R}^{B\times N_{\rm sq}\times N_{\rm H}},\\
    \tC&=s_{C}(\mathbf{\tH^{\rm h,ad}})\in \mathbb{R}^{B\times N_{\rm sq}\times N_{\rm H}},\\
    \tD&=\tau_{d}(s_{d}(\tH^{\rm h,ad})+\mathbf{b}^{\rm d})\in \mathbb{R}^{B\times N_{\rm sq}\times d_{\rm f}},
\end{aligned}\label{selective}
             \end{align}
where the $s_{B}(\mathbf{\tH^{\rm h,ad}})=\operatorname{FC}_{N_{\rm H}}(\mathbf{\tH^{\rm h,ad}})$, $s_{C}(\mathbf{\tH^{\rm h,ad}})=\operatorname{FC}_{N_{\rm H}}(\mathbf{\tH^{\rm h,ad}})$, $s_{d}(\mathbf{\tH^{\rm h,ad}})=\operatorname{Broadcast}_{d_{\rm f}}(\operatorname{FC}_{1}(\mathbf{\tH^{\rm h,ad}}))$, $\operatorname{Broadcast}_{x}$ is a broadcast projection to dimension $x$, $\tau_{d}$ denotes the softplus function, and $\mathbf{b}^{\rm d}\in \mathbb{R}^{d_{\rm f}\times 1}$ denotes a learnable vector. 
Building upon these input-dependent parameters, the discrete SSM in selective SSM \eqref{sssm_d} can be reformulated in the standard form
	\begin{align}
		\begin{aligned}
			\mathbf{s}_{b,t,f}&=  \bar{\mathbf{A}}_{b,t,f}\mathbf{s}_{b,t-1,f}+ \bar{\mathbf{b}}_{b,t,f} h_{b,t,f},\\
	\bar{h}_{b,t,f}&=\mathbf{c}_{b,t}^{T}\mathbf{s}_{b,t,f},
		\end{aligned}
        \label{rnn_s}
	\end{align}
where we define  $\bar{\mathbf{A}}_{b,t,f}=f_A(d_{b,t,f},\mathbf{A}_{f})$ and $\bar{\mathbf{b}}_{b,t,f}=f_B(d_{b,t,f},\mathbf{A}_{f},\mathbf{b}_{b,t})$ based on \eqref{dis}, $\mathbf{b}_{b,t} = \mathbf{b}_{bt:} \in \mathbb{R}^{N_{\rm H}\times 1}$ denotes the $(b,t)$-th fiber of $\tB$, and $d_{b,t,f}$ denotes the $(b,t,f)$-th element of  $\tD$.
The serial computation in \eqref{rnn_s} is well-suited for streaming inference but inefficient for gradient propagation. To mitigate this limitation, \eqref{rnn_s} can be reformulated into a CNN-like expression as follows, which is training-friendly and computationally parallelizable:
	\begin{align}
		\begin{aligned}
\textstyle\bar{h}_{b,t,f}=\sum_{j=1}^{t}\mathbf{c}_{b,j}^T\bar{\mathbf{A}}^{\rm p}_{b,j,f}\bar{\mathbf{b}}_{b,j,f}h_{b,j,f},\label{cov_s}
		\end{aligned}
	\end{align}
where the $\bar{\mathbf{A}}^{\rm p}_{b,j,f}=\prod_{k=j+1}^{t}\bar{\mathbf{A}}_{b,k,f}$.  It is noted that the conventional SSM becomes a special case of the selective SSM when the parameters of selective SSM become input-independent and time-invariant. 

In summary, the Mamba is trained on input tensor $\tH^{\rm h,ad}$, where the optimal tensors $\bar{\tA},\bar{\tB}$ and $\tC$ are obtained according to \eqref{selective} and \eqref{dis}, enabling the selective SSM \eqref{sssm} to perform output mapping based on the mechanism \eqref{rnn_s} or \eqref{cov_s}. This adaptation makes Mamba a time-variant and input-dependent model, enhancing its ability to capture and represent multi-domain features effectively.

			\begin{algorithm}[t]
			\caption{Training algorithm of the proposed network} 
			\label{Modeltraining}
			\begin{algorithmic}[1]
				\REQUIRE {Training dataset $ \mathcal{D}_{\rm train}$, network $f_{\rm TMCFN}$, $f_{\rm JUDCEN}$, and $f_{\rm MCPN}$.}
				\ENSURE {Trained network $f_{\rm HASCAN}$.}
				\STATE  {Randomly initialize $f_{\rm TMCFN}$, $f_{\rm JUDCEN}$, and $f_{\rm MCPN}$.}
                \STATE {\textbf{Phase I: TMCFN Pre-training}}
				\STATE {Train $f_{\rm TMCFN}$ under  $L^{\prime}_{\rm MSE}$ with ${\hat{\tG}^{\rm d}}$ in $ \mathcal{D}_{\rm train}$.}
                \STATE {\textbf{Phase II: JUDCEN and MCPN Pre-training}}
                \STATE {With the perfect CSI feedback assumption, train $f_{\rm JUDCEN}$ and $f_{\rm MCPN}$ under 
                $L_{\rm MSE}$ with ${\hat{\tG}^{\rm u}}$ and ${\hat{\tG}^{\rm d}}$ in $ \mathcal{D}_{\rm train}$.}
                \STATE {\textbf{Phase III: Joint End-to-End Fine-tuning}}
                \STATE {Initialize  $f_{\rm HASCAN}$ based on the optimal models $f_{\rm TMCFN}$, $f_{\rm JUDCEN}$ and $f_{\rm MCPN}$.}
                \STATE {Train $f_{\rm HASCAN}$ under $L_{\rm MSE}$ or $L_{\rm CosSim}$ with ${\hat{\tG}^{\rm u}}$ and ${\hat{\tG}^{\rm d}}$ in $ \mathcal{D}_{\rm train}$.}
            \end{algorithmic}
		\end{algorithm}	
\subsection{Training Algorithm}
	The multi-module architecture of HASCAN poses challenges for effective training. To enhance channel prediction performance, we propose a staged training strategy detailed in Algorithm \ref{Modeltraining}. Our training dataset, $ \mathcal{D}_{\rm train} =\left\{\hat{\tG}^{{\rm d}(i)},\hat{\tG}^{{\rm u}(i)},{\tG^{{\rm d}(i)}},\tH^{{\rm dt}(i)}\right\}_{i=1}^{N_{\rm train}}$, comprises $ N_{\rm train} $ samples with mixed SNRs. We adopt the adaptive moment estimation (Adam) optimizer to train the HASCAN  on $ \mathcal{D}_{\rm train} $ by PyTorch with four NVIDIA GeForce GTX 4090 GPUs.

    The training proceeds in three phases:
    \begin{itemize}
        \item \textbf{TMCFN Pre-training}: The TMCFN for CSI feedback is initially trained to minimize the mean squared error (MSE) loss between the initial DL estimated channel feature and its recovered counterpart: 
        \begin{align} 
        \begin{aligned} 
         L^{\prime}_{\rm MSE}&\textstyle= \sum_{i=1}^{N_{\rm train}}\|{\hat{\tG}^{{\rm fb} (i)}}-{\tG^{{\rm d}(i)}} \|_2^2.
        \end{aligned} 
        \end{align}
        \item \textbf{JUDCEN and MCPN Pre-training}: Subsequently, the JUDCEN and MCPN are trained to reconstruct and predict the spatial-frequency domain channels, with the initial UL and DL channel features as inputs. During this phase, the CSI feedback module is temporarily excluded, i.e., perfect CSI feedback is assumed, and the objective is to minimize the MSE of predicted and ground-truth channels, given by
        \begin{align}
            \begin{aligned}
               L_{\rm MSE}&\textstyle= \sum_{i=1}^{N_{\rm train}}\|\hat{\tH}^{{\rm dt}(i)}-\tH^{{\rm dt}(i)} \|_2^2,
        \end{aligned}
        \end{align}
        \item \textbf{Joint End-to-End Fine-tuning}: Finally, the overall joint training refines the pre-trained TMCFN, JUDCEN, and MCPN to maximize DL transmission efficiency. This final phase considers two channel estimation criteria: MSE minimization, as defined in the previous phase, and negative cosine similarity defined by
        \begin{align}
            \begin{aligned}
            L_{\rm CosSim}&\textstyle= -\sum_{i=1}^{N_{\rm train}}\frac{\hat{\tH}^{{\rm dt}(i)}\cdot\tH^{{\rm dt}(i)} }{\|\hat{\tH}^{{\rm dt}(i)} \|_2\|\tH^{{\rm dt}(i)} \|_2}.
            \end{aligned}
        \end{align}
    \end{itemize}
    \begin{table}[!tp]
	\centering
	\caption{Summary of Simulation Setting}
	\label{tab:simulation_setting}
	\scalebox{0.88}{
		\begin{tabular}{|c|c|c|c|}
			\hline
			Parameters & Values & Parameters & Values \\
			\hline
			\multicolumn{4}{|c|}{System Configuration} \\
			\hline
			Transmit antenna $ N_{\rm Tx} $ &  64     &Bandwidth  &100MHz \\
			Receive antenna $N_{\rm Rx}$& 8  &  Center frequency $ f_{\rm c} $      &  12GHz\\
			Fullband sounding $K $     &   2         &  Transmit RB $ N_{\rm RB} $ &52  \\ 
			Number of subbands $L$      &   4&  Active subcarrier   $ N_{\rm c} $        &  624\\
			DL transmission slots  $ N_{\rm tr}$     &  2&  CSI-RS ports $M_{\rm c}$      &   32\\
		DL training slots $ N_{\rm dt} $       &   4&SRS ports  $ M_{\rm s}$&4\\
		\begin{tabular}[c]{@{}c@{}}Time interval \\between SRSs\end{tabular}	     &   10ms&Transmission combs      &  4\\
		\begin{tabular}[c]{@{}c@{}}Time interval \\between CSI-RSs\end{tabular}	     &   5ms	& Channel model &\begin{tabular}[c]{@{}c@{}}38.901 UMa \\NLOS\end{tabular}\\
			\hline
			\multicolumn{4}{|c|}{Simulation Hyper-Parameter Settings} \\
			\hline
			Inner dimension	$ d_{\rm f}  $	& 512	& 	Multi-head $ N_{\rm m} $              &                 2       \\
			SABs	$ N_{\rm SA} $	& 6	&  	Mamba block	$ N_{\rm MA} $    & 6       \\
			Hidden dimension	$ N_{\rm H} $    & 128	&  Dropout probability	$ p$    & 0.1      \\
			Learning rate  	&  $ 6\times10^{-5} $	&  	Batch size  &              512      \\
			\hline
		\end{tabular}
	}
	\label{para}
\end{table}
	\section{Numerical Results}\label{Section 5}
In this section, we demonstrate the superiority of the proposed framework and networks through numerical simulations.
Specifically, we utilize QuaDRiGa \cite{jaeckel2014quadriga} to construct the simulation environment and generate the corresponding channels, which is capable of generating time-varying massive MIMO-OFDM channels that comply with the 3GPP New Radio (NR) specifications \cite{3gpp.38.901}. Additionally, 40,000 samples are generated with mixed SNRs, which are divided into 32,000 ($N_{\rm train}$), 4,000, and 4,000 samples for training, validation, and testing, respectively. The system configuration \cite{3gpp.38.211} and simulation hyper-parameter settings are shown in Table \ref{para}. 
To demonstrate the superiority of the proposed algorithm in various scenarios, we take the following channel estimation approaches as benchmarks:
	\begin{itemize}	
		\item \textbf{Ratio-Type II \cite{lee2021downlink}:}
		Ratio-Type II utilizes the Type II codebook for DL feedback and employs the ratio technique to extract the channel ratios at the receiver side.
		\item \textbf{MMSE-Type II \cite{noh2006low}:} MMSE-Type II employs the Type II codebook for DL feedback and leverages it to compute the correlation matrix at the receiver side.	
        \item \label{LS}\textbf{LCE  \cite{li2000optimum}:} 
		Linear channel extrapolation (LCE) performs linear extrapolation in the spatial and temporal domain to achieve full channel estimation and channel prediction.	
        \item \textbf{KDD-SFCEN + TUDCEN \cite{zhou2025low}:} 
		KDD-SFCEN utilizes self-attention for element-wise channel extrapolation in the spatial and frequency domains, while TUDCEN adopts generative AI for temporal-domain extrapolation.	
        \item \textbf{DACEN \cite{zhou2023pay}:}  
		The DACEN network, built upon a dual-attention mechanism, directly takes the received pilot signals as input and is trained to minimize the reconstruction error in the spatial-delay domain.
        \item \textbf{HASCAN:}  
		The network proposed in Section \ref{Section 4}, comprising  TMCFN, JUDCEN and MCPN.
	\end{itemize}

Among these benchmarks, DACEN exploits only DL pilots, LCE and KDD-SFCEN + TUDCEN utilize only UL pilots, while ratio-Type II, MMSE-Type II, and the proposed method leverage both UL and DL pilots.

\begin{figure*}[t]
  \centering
  \begin{minipage}{0.32\textwidth}
    \centering
    \includegraphics[width = \linewidth]{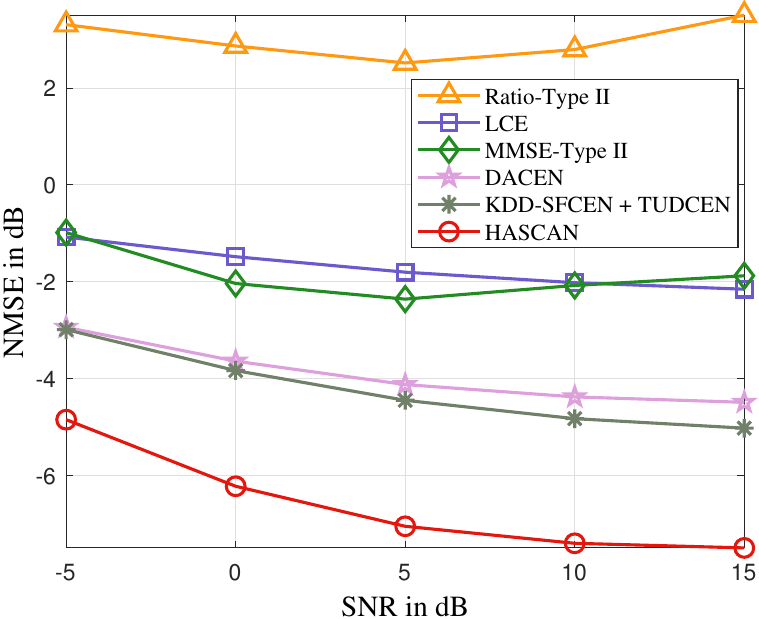}
    \caption{NMSE vs. SNR, with $N_{\rm bit}=400$ for CSI feedback.}
    \label{snrw}
  \end{minipage}\hfill
  \begin{minipage}{0.32\textwidth}
    \centering
\includegraphics[width = \linewidth]{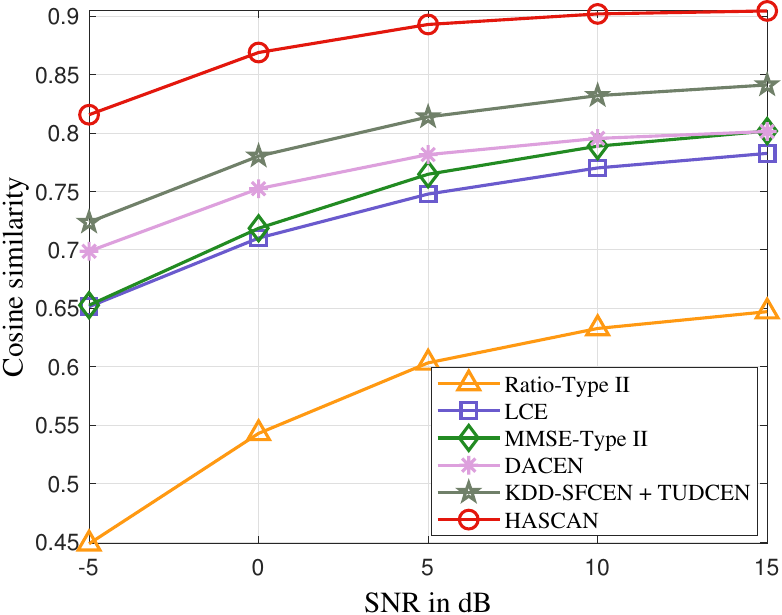}
    \caption{Cosine similarity vs. SNR, with $N_{\rm bit}=400$ for CSI feedback.}
    \label{cosw}
  \end{minipage}\hfill
  \begin{minipage}{0.32\textwidth}
    \centering
\includegraphics[width = \linewidth]{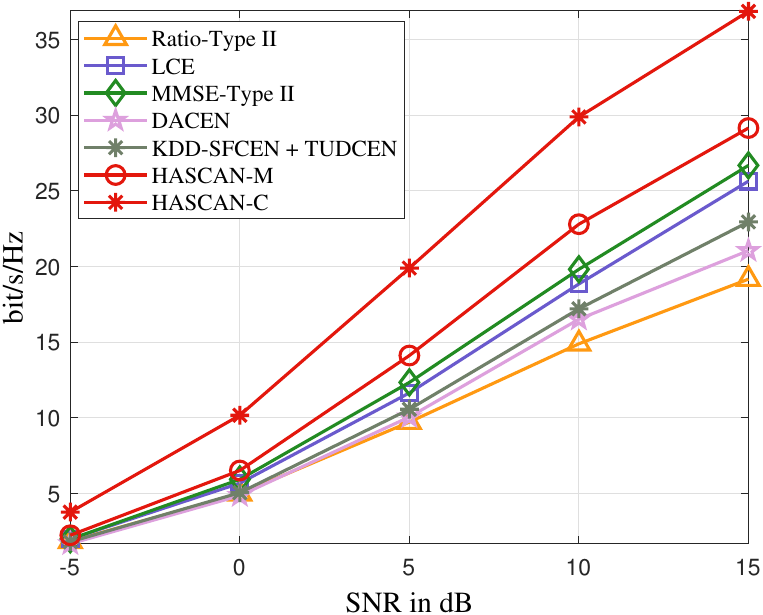}
    \caption{Spectral efficiency vs. SNR, with $N_{\rm bit}=400$ for CSI feedback and EZF for precoding.}
    \label{ezf}
  \end{minipage}
  \vspace{-5mm}
\end{figure*}

\subsection{CSI Acquisition Performance}
 \figref{snrw} illustrates the performance of all compared schemes in scenarios of various SNRs with $N_{\rm bit}=400$ for CSI feedback, where the data-driven approaches are trained on the mixed SNR dataset, and all methods are tested under different SNRs. 
 Notably, the DL channels are predicted for $N_{\rm tr}=2$ future transmission slots, with each slot spaced by $S\Delta t$ following the last SRS subband sounding.
 {Obviously, the performance of LCE is relatively poor due to the inability to exploit  the correlation among different antennas and slots. MMSE-Type II achieves a certain performance gain by utilizing DL information from Type II codebook feedback to obtain the antenna correlation matrix, which assists the extrapolation process. In contrast, although the ratio technique also attempts to exploit the feedback information, it only considers a simple single-carrier system and thus performs poorly in this scenario, where the precoding matrix indicator (PMI) codebook and SRS signal dimensions are mismatched.} Additionally, data-driven NN-based methods are capable of extracting additional information from datasets, such as KDD-SFCEN + TUDCEN, DACEN, and HASCAN, have shown significant improvements in performance. Specifically,  the DACEN enhances channel estimation  by leveraging dual-attention mechanisms to process the spatial-delay domain channel. The KDD-SFCEN combined with TUDCEN utilizes Transformer encoders in the spatial domain, while a generative Transformer performs recursive  temporal channel extrapolation. Furthermore, HASCAN employs a spatial-frequency attention mechanism in the spatial-frequency domain to effectively integrate UL and DL channels. In the angle-delay domain, the Mamba architecture is utilized for temporal extrapolation, leveraging its strength in handling long sequences. Obviously, relying solely on either UL pilot training or DL quantized feedback cannot yield optimal prediction performance. However, the proposed framework enables the HASCAN to jointly exploit both UL and DL signaling, and outperforms the DL-only method DACEN and the UL-only method KDD-SFCEN + TUDCEN  with improvements of $ 3.00 $dB and $ 2.47 $dB when $\text{SNR}=15$dB, respectively. As a result, the proposed network achieves significant performance gains over baseline approaches in FR3 massive MIMO scenarios, which clearly demonstrates the superiority of the proposed framework. 

   Table \ref{bits} shows the performance of all compared schemes under varying feedback overhead when  $\text{SNR} = 5  \text{dB} $. Specifically, it can be observed that the performance of the proposed method gradually improves as the feedback overhead increases. Moreover, LCE, DACEN, and KDD-SFCEN + TUDCEN exhibit constant performance as they assume perfect feedback or neglect CSI feedback. The ratio and MMSE approaches show limited performance variation with  respect to feedback bits, which may be attributed to their inefficient utilization of PMI.
  Since the HASCAN leverages the Transformer-based autoencoder and the upgraded TFFL to jointly exploit the channel sparsity in the angle-delay domain, it exhibits only limited degradation with fewer feedback bits. 
  Additionally,  the proposed training Algorithm \ref{Modeltraining} further ensures minimal performance degradation in joint CSI feedback and channel reconstruction. 
 Notably, the proposed method nearly achieves a temporal prediction performance of $-7$ dB with only $200$ bits in a large-scale system, underscoring its effectiveness and superiority. 

Furthermore, to evaluate the robustness of the considered approaches across the entire FR3 band, we compare the NMSE performance at four operating frequencies when  $\text{SNR} = 5  \text{dB} $, as shown in Table \ref{centerf}.  It can be observed that the CSI acquisition accuracy of all methods decreases with increasing center frequency, since higher frequencies exacerbate channel time-variation. Notably, the proposed method consistently outperforms all baseline approaches and maintains a channel prediction accuracy above $-5 \text{dB}$, even at $f_{\rm c}=22\text{GHz}$. This further demonstrates the superiority and robustness of the proposed framework, showing its effectiveness throughout the full FR3 band.

\subsection{DL Transmission Performance}
To further investigate the performance of the proposed method in multi-user DL transmission, we additionally generate 4,000 channel samples comprising 125 distinct cells, each containing 16 UEs. We first compare the cosine similarity performance of all methods, followed by an evaluation of their spectral efficiency in DL transmission, where the precoding weights are obtained by eigen zero-forcing (EZF) \cite{sun2010eigen}.
	 \figref{cosw} illustrates the cosine similarity performance of all compared schemes in scenarios of various SNRs with $N_{\rm bit}=400$ for CSI feedback. Unlike previous scenarios, the Type II codebook feeds back eigenvectors, which preserve angle information more effectively. As a result, the MMSE-Type II method exhibits a greater advantage over the LCE method in terms of cosine similarity performance and can even surpass the AI-based method DACEN under high SNR conditions. Similarly, HASCAN consistently achieves the best cosine similarity performance across all SNR levels. 

	\figref{ezf} illustrates the spectral efficiency performance of all compared schemes under various SNRs  with $N_{\rm bit}=400$ for CSI feedback during DL transmission.  In the simulations, we further compare the performance of the proposed method under different loss functions. The model trained with MMSE loss is denoted as HASCAN-M, while the one trained with cosine similarity loss is denoted as HASCAN-C. As shown in  \figref{ezf}, although the ratio technique performs poorly in terms of NMSE, its angle-domain prediction is still relatively reasonable, which prevents its spectral efficiency from degrading drastically. This motivates our use of cosine similarity as a training loss in HASCAN. Obviously, HASCAN-M achieves the best spectral efficiency across all SNR levels. Moreover, HASCAN-C, trained with cosine similarity loss, further improves spectral efficiency by $ 26\% $ over HASCAN-M when  $\text{SNR} = 15  \text{dB} $, without introducing any additional computational complexity. These results demonstrate the superiority and flexibility of the proposed framework  for DL transmission.  
\subsection{Ablation Experiment}
\begin{table*}[htbp]
\centering
\begin{minipage}[t]{0.48\textwidth}
\centering
\caption{NMSE Vs. Feedback Overhead}
\label{bits}
\begin{tabular}{c|c|c|c} 
\hline
\multirow{2}{*}{Method} & \multicolumn{3}{c}{NMSE(dB) vs. $N_{\rm bit}$} \\ 
\cline{2-4}
 & $200$ & $300$ & $400$ \\ 
\hline
Ratio-Type II & $2.50$ & $2.51$ & $2.52$ \\ 
\hline
LCE & $-1.80$ & $-1.80$ & $-1.80$ \\ 
\hline
MMSE-Type II & $-2.34$ & $-2.35$ & $-2.36$ \\ 
\hline
DACEN & $-4.12$ & $-4.12$ & $-4.12$ \\ 
\hline
KDD-SFCEN + TUDCEN & $-4.44$ & $-4.44$ & $-4.44$ \\ 
\hline
HASCAN (This work) & $\mathbf{-6.80}$ & $\mathbf{-6.87}$ & $\mathbf{-7.04}$ \\ 
\hline
\end{tabular}
\end{minipage}
\hfill
\begin{minipage}[t]{0.48\textwidth}
\centering
\caption{NMSE Vs. Center Frequency}
\label{centerf}
\begin{tabular}{c|c|c|c|c} 
\hline
\multirow{2}{*}{Method} & \multicolumn{4}{c}{NMSE(dB) vs. $f_{\rm c}$} \\ 
\cline{2-5}
& $7\text{GHz}$ & $12\text{GHz}$ & $17\text{GHz}$ & $22\text{GHz}$ \\ 
\hline
Ratio-Type II & $1.41$ & $2.52$ & $3.96$ & $5.10$ \\ 
\hline
LCE & $-3.33$ & $-1.80$ & $0.13$ & $1.79$ \\ 
\hline
MMSE-Type II & $-3.56$ & $-2.36$ & $-0.50$ & $1.20$ \\ 
\hline
DACEN & $-5.03$ & $-4.12$ & $-3.30$ & $-2.89$ \\ 
\hline
KDD-SFCEN + TUDCEN & $-4.87$ & $-4.44$ & $-3.78$ & $-3.06$ \\ 
\hline
HASCAN (This work) & $\mathbf{-7.80}$ & $\mathbf{-7.04}$ & $\mathbf{-6.25}$ & $\mathbf{-5.37}$ \\ 
\hline
\end{tabular}
\end{minipage}
\vspace{-5mm}
\end{table*}
\begin{table*}
\centering
\caption{Computational Complexity Comparison and Ablation Experiment of NN-based Approaches}
\label{ab}
\begin{tabular}{c|c|c|c|c|c|c|c} 
\hline
\multirow{2}{*}{Method} & \multicolumn{4}{c|}{Module}                                              & \multicolumn{2}{c|}{Computational Complexity}                       & \multirow{2}{*}{NMSE(dB)}  \\ 
\cline{2-7}
                        & DL estimation & CSI feedback & UL estimation      & Joint reconstruction & FLOPs                                     & Params(MB)              &                            \\ 
\hline
HASCAN (This work)                 & HASCAN        & HASCAN       & HASCAN             & HASCAN               & $\mathbf{8.21}\textbf{G}$                            & $\mathbf{69.88}$          & $\mathbf{-6.22}$             \\ 
\hline
DACEN-J                 & DACEN         & HASCAN       & HASCAN             & HASCAN               & $13.32\text{G}$                           & $70.86 $                  & $-5.17$                      \\ 
\hline
KDD-J    & HASCAN        & HASCAN       & KDD & HASCAN               & $8.35\text{G}$                                   & $104.45$                   & $-4.57  $                    \\ 
\hline
HASCAN-D                & HASCAN        & HASCAN       & ——                 & ——                   & $\mathbf{8.16}\textbf{G}$                   &$ \mathbf{69.75}$ & $\mathbf{-3.91} $            \\
\hline
DACEN                   & DACEN         & HASCAN       & ——                 & ——                   & $12.61\text{G}  $                                & $70.53$                   & $-3.64 $                     \\ 
\hline
HASCAN-U                & ——            & ——           & HASCAN             & ——                   & $\mathbf{7.86}\textbf{G}$ & $\mathbf{42.22}$ & $\mathbf{-4.58} $            \\
\hline
KDD      & ——            & ——           & KDD & ——                   & $7.94\text{G} $                                  & $76.52 $                  & $-3.83$                      \\ 
\hline
\end{tabular}
\end{table*}
To further evaluate the superiority of the designed networks, we conduct an ablation experiment. Specifically, we define HASCAN-U and HASCAN-D as variants that use only UL pilot training or only DL quantized feedback, respectively. As shown in Table \ref{ab},  KDD denotes KDD-SFCEN + TUDCEN  for brevity, KDD-J and DACEN-J represent the KDD-SFCEN + TUDCEN and DACEN networks, respectively, which perform channel reconstruction by jointly exploiting UL and DL pilots within the proposed framework.
When $\text{SNR}=0$dB and $N_{\rm bit}=400$, Table \ref{ab}  exhibits the CSI acquisition performance of the proposed method and other NN-based approaches with joint UL and DL pilot or with only UL/DL pilot. 
As indicated in the final four rows in Table \ref{ab}, HASCAN-U also outperforms  the  KDD-SFCEN + TUDCEN with only SRS inputs, while HASCAN-D outperform  the  DACEN with only CSI-RS inputs, validating the effectiveness of the proposed network design. 
Furthermore, within the proposed framework, additional DL/UL pilots can be provided to networks KDD-SFCEN + TUDCEN and DACEN by HASCAN to enable joint CSI acquisition.
Evidently, KDD-J and DACEN-J both achieve corresponding performance gains, yet their performance falls short of the proposed method, further demonstrating the superiority of the designed network and framework.

\subsection{Complexity}
In addition, we compare the proposed algorithms in terms of the number of floating-point operations (FLOPs) and the total number of network parameters, which are summarized in Table \ref{ab}.
Benefiting from the SSM structure that models sequence dependencies via a linear convolution formulation,  Mamba eliminates the need for explicit pairwise interaction computation across all positions as required by Transformers. Moreover, the parameters of the SSM kernel, such as $\bar{\mathbf{A}}_{b,t,f}$, $\bar{\mathbf{b}}_{b,t,f}$ and $\mathbf{c}_{b,t}$ can be expressed in low-rank or structured forms, further reducing parameter requirements. Consequently, the proposed network maintains the lowest parameter count among all baseline methods. Meanwhile, it can be observed that the proposed network, owing to the linear complexity of the SSM architecture, achieves significantly lower FLOPs than other methods.  Compared to the DACEN network based on the Transformer architecture with quadratic complexity, the proposed method achieves a roughly $ 38\% $ reduction in FLOPs.

\section{Conclusion}\label{Section 6}
In this paper, we proposed a joint UL and DL transmission framework and designed dedicated neural networks for each key component, which addressed the challenges posed by the emerging and promising FR3 band. Specifically, the TMCFN was developed to enhance sparse-domain CSI feedback using Transformer and MLP architectures. Furthermore, the proposed  spatial-frequency-attention-based JUDCEN and the SSM-based MCPN, enabled accurate full-dimensional channel reconstruction and prediction by jointly utilizing UL and DL pilot signals.  Extensive simulations demonstrated the proposed framework and networks consistently outperformed benchmark approaches in terms of NMSE, cosine similarity, and spectral efficiency. Moreover, the proposed framework and networks exhibited robustness under different SNR levels and operating frequencies, while achieving reduced feedback overhead and computational complexity.


	
	
	%
	\bibliographystyle{IEEEtran}
	\bibliography{IEEEabrv,IEEEfull}

@STRING{IEEE_J_VT         = "{IEEE} Trans. Veh. Technol."}

@STRING{IEEE_J_STSP       = "{IEEE} J. Sel. Topics Signal Process."}

@STRING{IEEE_J_SP         = "{IEEE} Trans. Signal Process."}

@STRING{IEEE_J_JSAC       = "{IEEE} J. Sel. Areas Commun."}

@STRING{IEEE_J_COM        = "{IEEE} Trans. Commun."}

@STRING{IEEE_J_WCOM       = "{IEEE} Trans. Wireless Commun."}

@STRING{IEEE_J_WCOML      = "{IEEE} Wireless Commun. Lett."}

@STRING{IEEE_J_IOT        = "{IEEE} Internet Things J."}

@STRING{IEEE_J_AP         = "{IEEE} Trans. Antennas Propag."}

@STRING{IEEE_M_COM        = "{IEEE} Commun. Mag."}

@STRING{IEEE_O_CSTO       = "{IEEE} Commun. Surveys Tuts."}

@IEEEtranBSTCTL{IEEEexample:BSTcontrol, 
	CTLuse_forced_etal    = "yes",
	CTLmax_names_forced_etal  = "3",
	CTLnames_show_etal   = "2",
}

@article{giordani2020toward,
  title={{Toward 6G networks: Use cases and technologies}},
  author={Giordani, Marco and Polese, Michele and Mezzavilla, Marco and Rangan, Sundeep and Zorzi, Michele},
  journal=IEEE_M_COM,
  volume={58},
  number={3},
  pages={55--61},
  year={Dec. 2020},
  publisher={IEEE}
}

@Article{naqvi20215g,
  author    = {Naqvi, Syed Hassan Raza and Ho, Pin Han and Peng, Limei},
  journal   = {J. Commun. Netw.},
  title     = {{5G NR mmWave indoor coverage with massive antenna system}},
  year      = {Feb. 2021},
  number    = {1},
  pages     = {1--11},
  volume    = {23},
  publisher = {KICS},
}

@misc{2022Samsung,
  howpublished = {Samsung. (May 2022). {\it 6G Spectrum: Expanding the Frontier.} [Online]. Available: \url{https://cdn.codeground.org/nsr/downloads/researchareas/2022May 6G Spectrum.pdf}}
}

@misc{2022je,
  howpublished = {J. E. Smee. (Feb. 2022). {\it 10 Innovation Areas for 5G Advanced and Beyond [Video].} [Online]. Available: \url{https://www.qualcomm.com/news/onq/2022/02/10-innovation-areas-5g-advanced-and-beyond}}
}

@InProceedings{2023WRC,
  author    = {ITU-R},
  booktitle = {{World Radiocommunication Conference 2023 (WRC-23)}},
  title     = {{Provisional final acts}},
  year      = {Nov. 2023},
  url       = {https://www.itu.int/dms_pub/itu-r/opb/act/R-ACT-WRC.15-2023-PDF-E.pdf},
  address   = {Dubai, United Arab Emirates},
}

@article{hou2025tensor,
  title={{Tensor-structured Bayesian channel prediction for upper mid-band XL-MIMO systems}},
  author={Hou, Hongwei and Wang, Yafei and Yi, Xinping and Wang, Wenjin and Slock, Dirk and Jin, Shi},
  journal={arXiv:2508.08491},
  year={Aug. 2025},
  url={https://arxiv.org/abs/2508.08491},
}

@article{wang2022weighted,
  title={{Weighted MMSE precoding for constructive interference region}},
  author={Wang, Yafei and Wang, Wenjin and You, Li and Tsinos, Christos G and Jin, Shi},
  journal=IEEE_J_WCOML,
  volume={11},
  number={12},
  pages={2605--2609},
  year={Oct. 2022},
  publisher={IEEE}
}

@article{wang2025statistical,
  title={{Statistical CSI-based distributed precoding design for OFDM-cooperative multi-satellite systems}},
  author={Wang, Yafei and Ha, Vu Nguyen and Ntontin, Konstantinos and Yan, Hong and Wang, Wenjin and Chatzinotas, Symeon and Ottersten, Bj{\"o}rn},
  journal={arXiv:2505.08038},
  year={May 2025},
  url={https://arxiv.org/abs/2505.08038},
}

@ARTICLE{11049893,
  author={Wang, Yafei and Hou, Hongwei and Yi, Xinping and Wang, Wenjin and Jin, Shi},
  journal=IEEE_J_WCOM, 
  title={{Towards unified AI models for MU-MIMO communications: A tensor equivariance framework}}, 
  year={Early Access, 2025},
  volume={},
  number={},
  doi={10.1109/TWC.2025.3580321}}

@article{WANG2025,
title = {{Toward mobile satellite internet: The fundamental limitation of wireless transmission and enabling technologies}},
journal = {Engineering},
year = {Jul. 2025},
issn = {2095-8099},
doi = {https://doi.org/10.1016/j.eng.2025.07.007},
url = {https://www.sciencedirect.com/science/article/pii/S2095809925003698},
author = {Wenjin Wang and Yiming Zhu and Yafei Wang and Rui Ding and Symeon Chatzinotas}
}

@TechReport{3gpp.38.211,
  author      = {{3rd Generation Partnership Project (3GPP)}},
  institution = {{3GPP}},
  title       = {{Physical channels and modulation}},
  year        = {Jun. 2021},
  number      = {38.211,{V.16.6.0}},
  type        = {TS},
}

@article{miao2023sub,
  title={{Sub-6 GHz to mmWave for 5G-advanced and beyond: Channel measurements, characteristics and impact on system performance}},
  author={Miao, Haiyang and Zhang, Jianhua and Tang, Pan and Tian, Lei and Zhao, Xinyu and Guo, Bolun and Liu, Guangyi},
  journal=IEEE_J_JSAC,
  volume={41},
  number={6},
  pages={1945--1960},
  year={May 2023},
  publisher={IEEE}
}

@article{zheng2022survey,
  title={{A survey on channel estimation and practical passive beamforming design for intelligent reflecting surface aided wireless communications}},
  author={Zheng, Beixiong and You, Changsheng and Mei, Weidong and Zhang, Rui},
  journal=IEEE_O_CSTO,
  volume={24},
  number={2},
  pages={1035--1071},
  year={Feb. 2022},
  publisher={IEEE}
}

@InProceedings{li2000optimum,
  author    = {Li, Ye},
  booktitle = {Proc. IEEE Global Telecommun. Conf. (GLOBECOM)},
  title     = {{Optimum training sequences for OFDM systems with multiple transmit antennas}},
  address = {San Francisco, CA, USA},
  year      = {2000},
  month     = {Nov.},
  pages     = {1478--1482},
  volume    = {3},
}

@article{noh2006low,
  title={{Low complexity LMMSE channel estimation for OFDM}},
  author={Noh, Minseok and Lee, Yusung and Park, Hyuncheol},
  journal={IEE Proc., Commun.},
  volume={153},
  number={5},
  pages={645--650},
  year={Oct. 2006},
  publisher={IET}
}

@article{zhu2024joint,
  title={{Joint channel estimation and prediction for massive MIMO with frequency hopping sounding}},
  author={Zhu, Yiming and Zhuang, Jiawei and Sun, Gangle and Hou, Hongwei and You, Li and Wang, Wenjin},
  journal=IEEE_J_COM,
  year={Dec. 2024},
  volume={73},
  number={7},
  pages={5139--5154},
  publisher={IEEE}
}

@Article{dong2019deep,
  author    = {Dong, Peihao and Zhang, Hua and Li, Geoffrey Ye and Gaspar, Ivan Simoes and NaderiAlizadeh, Navid},
  journal   = IEEE_J_STSP,
  title     = {Deep {CNN}-based channel estimation for mm{W}ave massive {MIMO} systems},
  year      = {2019},
  month     = {Sep.},
  number    = {5},
  pages     = {989--1000},
  volume    = {13},
  publisher = {IEEE},
}

@Article{lin2021deep,
  author    = {Lin, Bo and Gao, Feifei and Zhang, Shun and Zhou, Ting and Alkhateeb, Ahmed},
  journal   = IEEE_J_WCOM,
  title     = {{Deep learning-based antenna selection and CSI extrapolation in massive MIMO systems}},
  year      = {Jun. 2021},
  number    = {11},
  pages     = {7669--7681},
  volume    = {20},
  publisher = {IEEE},
}

@Article{lee2016channel,
  author    = {Lee, Junho and Gil, Gye-Tae and Lee, Yong H},
  journal   = IEEE_J_COM,
  title     = {Channel estimation via orthogonal matching pursuit for hybrid {MIMO} systems in millimeter wave communications},
  year      = {2016},
  month     = {Jun.},
  number    = {6},
  pages     = {2370--2386},
  volume    = {64},
  publisher = {IEEE},
}

@article{wu2021hybrid,
  title={{Hybrid channel estimation for UPA-assisted millimeter-wave massive MIMO IoT systems}},
  author={Wu, Xianda and Yang, Xi and Ma, Shaodan and Zhou, Binggui and Yang, Guanghua},
  journal=IEEE_J_IOT,
  volume={9},
  number={4},
  pages={2829--2842},
  year={Jul. 2021},
  publisher={IEEE}
}

@Article{zhou2023pay,
  author    = {Zhou, Binggui and Yang, Xi and Ma, Shaodan and Gao, Feifei and Yang, Guanghua},
  journal   = IEEE_J_WCOM,
  title     = {{Pay less but get more: A dual-attention-based channel estimation network for massive MIMO systems with low-density pilots}},
  year      = {Nov. 2023},
  number    = {6},
  pages     = {6061--6076},
  volume    = {23},
  publisher = {IEEE},
}

@article{he2024angle,
  title={{Angle-delay domain hybrid model-driven and data-driven downlink CSI acquisition for FDD massive MIMO systems}},
  author={He, Xuan and Hou, Hongwei and Fang, Tianhao and Wang, Wenjin and Jin, Shi},
  journal=IEEE_J_VT,
  year={Sep. 2024},
  number    = {1},
  pages     = {1788--1793},
  volume    = {74},
  publisher={IEEE}
}

@article{zhuang2025extract,
  title={{Extract the best, discard the rest: CSI feedback with offline large AI models}},
  author={Zhuang, Jialin and Wang, Yafei and Hou, Hongwei and Han, Yu and Wang, Wenjin and Jin, Shi and Wang, Jiangzhou},
  journal={arXiv:2505.08566},
  year={May 2025},
  url={https://arxiv.org/abs/2505.08566},
}

@article{wen2018deep,
  title={{Deep learning for massive MIMO CSI feedback}},
  author={Wen, Chao-Kai and Shih, Wan-Ting and Jin, Shi},
  journal=IEEE_J_WCOML,
  volume={7},
  number={5},
  pages={748--751},
  year={Mar. 2018},
  publisher={IEEE}
}

@article{cui2022transnet,
  title={{TransNet: Full attention network for CSI feedback in FDD massive MIMO system}},
  author={Cui, Yaodong and Guo, Aihuang and Song, Chunlin},
  journal=IEEE_J_WCOML,
  volume={11},
  number={5},
  pages={903--907},
  year={Feb. 2022},
  publisher={IEEE}
}

@InProceedings{vaswani2017attention,
  author    = {Vaswani, Ashish and Shazeer, Noam and Parmar, Niki and Uszkoreit, Jakob and Jones, Llion and Gomez, Aidan N and Kaiser, {\L}ukasz and Polosukhin, Illia},
  booktitle = {Proc. Adv. Neural Inf. Process. Syst. (NIPS)},
  title     = {Attention is all you need},
  year      = {2017},
  month     = {Dec.},
  pages     = {5998--6008},
}

@article{hou2024tensor,
  title={{A tensor-structured approach to dynamic channel prediction for massive MIMO systems with temporal non-stationarity}},
  author={Hou, Hongwei and Wang, Yafei and Zhu, Yiming and Yi, Xinping and Wang, Wenjin and Slock, Dirk and Jin, Shi},
  journal=IEEE_J_WCOM,
  year={Early Access, 2025},
}

@InProceedings{wong2006wlc43,
  author       = {Wong, Ian C and Evans, Brian L},
  booktitle    = {Proc. IEEE Global Telecommun. Conf. (GLOBECOM)},
  title        = {{WLC43-5: low-complexity adaptive high-resolution channel prediction for OFDM systems}},
  address  ={San Francisco, CA, USA},
  year         = {Nov. 2006},
  pages        = {1--5},
}

@Article{duel2007fading,
  author    = {Duel-Hallen, Alexandra},
  journal   = {Proc. IEEE},
  title     = {{Fading channel prediction for mobile radio adaptive transmission systems}},
  year      = {Dec. 2007},
  number    = {12},
  pages     = {2299--2313},
  volume    = {95},
  publisher = {IEEE},
}

@Article{wu2021channel,
  author    = {Wu, Chi and Yi, Xinping and Zhu, Yiming and Wang, Wenjin and You, Li and Gao, Xiqi},
  journal   = IEEE_J_JSAC,
  title     = {{Channel prediction in high-mobility massive {MIMO}: From spatio-temporal autoregression to deep learning}},
  year      = {2021},
  month     = {Jul.},
  number    = {7},
  pages     = {1915--1930},
  volume    = {39},
  publisher = {IEEE},
}

@Article{lv2019channel,
  author    = {Lv, Changwei and Lin, Jia-Chin and Yang, Zhaocheng},
  journal   = {IEEE Access},
  title     = {{Channel prediction for millimeter wave MIMO-OFDM communications in rapidly time-varying frequency-selective fading channels}},
  year      = {Jan. 2019},
  pages     = {15183--15195},
  volume    = {7},
  publisher = {IEEE},
}

@Article{zhou2025low,
  author    = {Zhou, Binggui and Yang, Xi and Ma, Shaodan and Gao, Feifei and Yang, Guanghua},
  journal   = IEEE_J_WCOM,
  title     = {{Low-overhead channel estimation via 3D extrapolation for TDD mmWave massive MIMO systems under high-mobility scenarios}},
  year      = {Jan. 2025},
  number    = {4},
  pages     = {2797--2813},
  volume    = {24},
  publisher = {IEEE},
}

@Article{salim2011combining,
  author    = {Salim, Umer and Gesbert, David and Slock, Dirk},
  journal   = IEEE_J_SP,
  title     = {{Combining training and quantized feedback in multiantenna reciprocal channels}},
  year      = {Dec. 2011},
  number    = {3},
  pages     = {1383--1396},
  volume    = {60},
  publisher = {IEEE},
}

@Article{lee2021downlink,
  author    = {Lee, Hyeongtaek and Choi, Hyuckjin and Kim, Hwanjin and Kim, Sucheol and Jang, Chulhee and Choi, Yongyun and Choi, Junil},
  journal   = IEEE_J_WCOM,
  title     = {{Downlink channel reconstruction for spatial multiplexing in massive MIMO systems}},
  year      = {Apr. 2021},
  number    = {9},
  pages     = {6154--6166},
  volume    = {20},
  publisher = {IEEE},
}

@article{hou2024joint,
  title={{Joint beam alignment and Doppler estimation for fast time-varying wideband mmWave channels}},
  author={Hou, Hongwei and Wang, Yafei and Yi, Xinping and Wang, Wenjin and Jin, Shi},
  journal=IEEE_J_WCOM,
  volume={23},
  number={9},
  pages={10895--10910},
  year={Mar. 2024},
  publisher={IEEE}
}

@article{guo2022overview,
  title={{Overview of deep learning-based CSI feedback in massive MIMO systems}},
  author={Guo, Jiajia and Wen, Chao-Kai and Jin, Shi and Li, Geoffrey Ye},
  journal=IEEE_J_COM,
  volume={70},
  number={12},
  pages={8017--8045},
  year={Oct. 2022},
  publisher={IEEE}
}

@article{tolstikhin2021mlp,
  title={{MLP-mixer: An all-MLP architecture for vision}},
  author={Tolstikhin, Ilya O and Houlsby, Neil and Kolesnikov, Alexander and Beyer, Lucas and Zhai, Xiaohua and Unterthiner, Thomas and Yung, Jessica and Steiner, Andreas and Keysers, Daniel and Uszkoreit, Jakob and others},
  journal={Proc. Adv. Neural Inf. Process. Syst. (NIPS)},
  volume={34},
  pages={24261--24272},
  year={Dec. 2021}
}

@article{zhuang2024covnet,
  title={{CovNet: Covariance information-assisted CSI feedback for FDD massive MIMO systems}},
  author={Zhuang, Jialin and He, Xuan and Wang, Yafei and Liu, Jiale and Wang, Wenjin},
  journal=IEEE_J_WCOML,
  year={Dec. 2024},
  volume={14},
  number={3},
  pages={641--645},
  publisher={IEEE}
}

@InProceedings{woo2018cbam,
  author    = {Woo, Sanghyun and Park, Jongchan and Lee, Joon-Young and Kweon, In So},
  booktitle = {Proc. Eur. Conf. Comput. Vis. (ECCV)},
  title     = {C{BAM}: Convolutional block attention module},
  year      = {2018},
  address   = {Munich, Germany},
  month     = {Sep.},
  pages     = {3--19},
}

@article{dai2019transformer,
  title={{Transformer-XL: Attentive language models beyond a fixed-length context}},
  author={Dai, Zihang and Yang, Zhilin and Yang, Yiming and Carbonell, Jaime and Le, Quoc V and Salakhutdinov, Ruslan},
  journal={arXiv:1901.02860},
  year={Jan. 2019},
  url={https://arxiv.org/abs/1901.02860},
}

@inproceedings{
gu2024mamba,
title={{Mamba: Linear-time sequence modeling with selective state spaces}},
author={Albert Gu and Tri Dao},
booktitle={First Conference on Language Modeling (COLM)},
year={Jul. 2024},
address={{Philadelphia, PA, USA}},
url={https://openreview.net/forum?id=tEYskw1VY2}
}

@inproceedings{
fu2023hungry,
title={{Hungry hungry hippos: Towards language modeling with state space models}},
author={Daniel Y Fu and Tri Dao and Khaled Kamal Saab and Armin W Thomas and Atri Rudra and Christopher Re},
booktitle={Proc. Int. Conf. Learn. Representations (ICLR)},
address   = {Kigali, Rwanda},
year={Feb. 2023},
url={https://openreview.net/forum?id=COZDy0WYGg}
}

@inproceedings{hua2022transformer,
  title={{Transformer quality in linear time}},
  author={Hua, Weizhe and Dai, Zihang and Liu, Hanxiao and Le, Quoc},
  booktitle={Proc. Int. Conf. Mach. Learn.},
  pages={9099--9117},
  address = {Honolulu, HI, USA},
  year={Jul. 2022},
}

@inproceedings{
gu2022efficiently,
title={{Efficiently modeling long sequences with structured state spaces}},
author={Albert Gu and Karan Goel and Christopher Re},
booktitle={Proc. Int. Conf. Learn. Representations (ICLR)},
year={Jan. 2022},
url={https://openreview.net/forum?id=uYLFoz1vlAC}
}

@article{gu2022parameterization,
  title={{On the parameterization and initialization of diagonal state space models}},
  author={Gu, Albert and Goel, Karan and Gupta, Ankit and R{\'e}, Christopher},
  journal={Proc. Adv. Neural Inf. Process. Syst. (NIPS)},
  volume={35},
  pages={35971--35983},
  year={Nov. 2022}
}

@article{gu2020hippo,
  title={{HIPPO: Recurrent memory with optimal polynomial projections}},
  author={Gu, Albert and Dao, Tri and Ermon, Stefano and Rudra, Atri and R{\'e}, Christopher},
  journal={Proc. Adv. Neural Inf. Process. Syst. (NIPS)},
  volume={33},
  pages={1474--1487},
  year={Dec. 2020}
}

@inproceedings{
martin2018parallelizing,
title={{Parallelizing linear recurrent neural nets over sequence length}},
author={Eric Martin and Chris Cundy},
booktitle={Proc. Int. Conf. Learn. Representations (ICLR)},
address   = {Vancouver, BC, Canada},
year={Jan. 2018},
url={https://openreview.net/forum?id=HyUNwulC-},
}

@article{jaeckel2014quadriga,
  title={{QuaDRiGa: A 3-D multi-cell channel model with time evolution for enabling virtual field trials}},
  author={Jaeckel, Stephan and Raschkowski, Leszek and B{\"o}rner, Kai and Thiele, Lars},
  journal=IEEE_J_AP,
  volume={62},
  number={6},
  pages={3242--3256},
  year={Mar. 2014},
  publisher={IEEE}
}

@TechReport{3gpp.38.901,
  author      = {{3rd Generation Partnership Project (3GPP)}},
  institution = {{3GPP}},
  title       = {{Study on channel model for frequencies from 0.5 to 100 GHz}},
  year        = {2020},
  number      = {38.901,{V.16.1.0}},
  type        = {TR},
}

@article{sun2010eigen,
  title={{Eigen-based transceivers for the MIMO broadcast channel with semi-orthogonal user selection}},
  author={Sun, Liang and McKay, Matthew R},
  journal=IEEE_J_SP,
  volume={58},
  number={10},
  pages={5246--5261},
  year={Jun. 2010},
  publisher={IEEE}
}

@article{he2025sca,
title={{SCA-LLM: Spectral-attentive channel prediction with large language models in MIMO-OFDM}},
  author={He, Ke and He, Le and Fan, Lisheng and Lei, Xianfu and Vu, Thang X and Karagiannidis, George K and Chatzinotas, Symeon},
  journal={arXiv:2509.08139},
  year={Sep. 2025},
  url={https://arxiv.org/abs/2509.08139}
}

@article{hou2024beam,
  title={{Beam-delay domain channel estimation for mmWave XL-MIMO systems}},
  author={Hou, Hongwei and He, Xuan and Fang, Tianhao and Yi, Xinping and Wang, Wenjin and Jin, Shi},
  journal=IEEE_J_STSP,
  volume={18},
  number={4},
  pages={646--661},
  year={May 2024},
  publisher={IEEE}
}

@article{fang2023multi,
  title={{A multi-beam XL-MIMO testbed based on hybrid CPU-FPGA architecture}},
  author={Fang, Tianhao and Gao, Yangyang and Suo, Chaoju and Sun, Gangle and Chen, Pengyu and Xiao, Wei and Wang, Wenjin},
  journal={Electronics},
  volume={12},
  number={2},
  pages={380},
  year={Jan. 2023},
  publisher={MDPI}
}
\end{document}